\documentclass{aa}

\usepackage{graphicx}
\usepackage{txfonts}
\usepackage{lipsum}
\usepackage{subcaption}         
\usepackage{lscape}             
\usepackage{placeins}           

\usepackage{amsmath}
\usepackage{booktabs,array,makecell,tabularx}

\usepackage[colorlinks=true, allcolors=blue]{hyperref}

\begin{document}

   \title{You're Gonna Need a Bigger Core}

   \subtitle{Calibrating Massive Star Models against Galactic OB-type Stars}

%
%
%

   \author{Thibault Lechien\inst{1}\corrauth{lechien@mpa-garching.mpg.de}        
        \and Selma E. de Mink\inst{1}\email{SEdeMink@mpa-garching.mpg.de}
        \and Stephen Justham\inst{1}\email{sjustham@mpa-garching.mpg.de}
        \and Abel de Burgos\inst{2}\email{abel.deburgossierra@eso.org}
        \and Ruggero Valli\inst{1}\email{ruvalli@mpa-garching.mpg.de}
        \and Sergio Simón-Díaz\inst{3,4}\email{ssimon@iac.es}
        }

   \institute{Max Planck Institute for Astrophysics, Karl-Schwarzschild-Straße 1, 85748 Garching bei München, Germany \and European Southern Observatory, Alonso de Córdova 3107, Vitacura, Santiago, Chile \and Instituto de Astrofísica de Canarias, E-38200 La Laguna, Tenerife, Spain \and Departamento de Astrofísica, Universidad de La Laguna, E-38205 La Laguna, Tenerife, Spain}

   \date{Received July 25, 2026}


  \abstract
   {The evolution of massive stars above 8 M$_\odot$ depends critically on the amount of mixing above the convective core during the main sequence.
	However, current models typically extrapolate results from lower-mass stars, where constraints from asteroseismology and eclipsing binary systems are more readily available.
	A new opportunity to study the evolution of massive stars and their distribution in the Hertzsprung--Russell diagram arises by
	combining the IACOB spectroscopic sample of over 900 Galactic OB-type stars with Gaia distances.
	We use this homogeneously analyzed sample to place population-level constraints on main-sequence evolution.
	We analyze the data by forward modeling stellar evolution tracks with MESA and applying Bayesian inference.
	This enables us to 1) determine a physically-motivated, data-driven location of the terminal-age main sequence, 2) constrain convective boundary mixing and resulting core masses, and 3) provide a set of massive star models calibrated against modern data.
	We explore how boundary mixing depends on mass and find that it
	is well described by a constant overshooting parameter in the mass range of 12 to 40~M$_\odot$, with $\alpha_{\mathrm{ov}} = 0.33 \pm 0.02$, or $f_{\mathrm{ov}} = 0.028 \pm 0.003$ in the step and exponential overshooting schemes respectively.
	We find evidence against a continuation of the trend to increase with mass that is found at lower masses. Instead, the data does not exclude a decreasing trend at the high mass end.
	We find that the resulting helium core masses are 10 to 40\% larger than other commonly used overshooting prescriptions.
	Combining our findings with existing observational constraints for low- and intermediate-mass stars, we propose a new mass-dependent overshooting prescription for a wide range of masses.
	Our calibration and model set are particularly useful for population and spectral synthesis applications.
}

   \keywords{   Stars: massive --
                Stars: evolution -- Convection
               }

   \maketitle
   \nolinenumbers

\section{Introduction}\label{s: Introduction}

Massive stars play a central role in astrophysics by driving chemical enrichment through supernovae, producing compact remnants such as black holes, shaping binary and multiple-star evolution, and contributing strongly to the ionizing radiation budget of galaxies \citep[see the reviews by, e.g.][]{
	iben_single_1984,
	langer_presupernova_2012,
	eldridge_new_2022}.

During the life of a star, the core and structure above it affects the stellar  evolution in many ways, including changing its luminosity, effective temperature, radius, and lifetime \citep[e.g.][]{maeder_stellar_1976,stothers_stellar_1985,herwig_evolution_2000,langer_presupernova_2012}.
The final fate of the star is also affected: it changes which stars form black holes and how massive the remnant will be \citep[e.g.][]{heger_presupernova_2000,ugliano_progenitor-explosion_2012,farmer_variations_2016,schneider_bimodal_2023,temaj_convective-core_2024}.
In a binary system, it changes when, and therefore how mass transfer occurs, leading to drastically different final products \citep{andersen_new_1990,tanikawa_population_2021,agrawal_modelling_2023,iorio_compact_2023}.

Despite its importance, mixing at the core boundary and the resulting core masses remain a key uncertainty in the evolution of massive stars
\citep[e.g.][]{roxburgh_note_1965,saslaw_overshooting_1965,maeder_grids_1987,zahn_convective_1991,kaiser_relative_2020}.
When convective core boundary mixing -- and consequently, stellar evolutionary tracks -- is not correctly modeled, this leads to large systematic errors in observable predictions, such as the predicted stellar spectra and the integrated light of galaxies \citep{leitherer_starburst99_1999,eldridge_binary_2017}.

Physically, several hydrodynamic processes are expected to drive mixing at the convective boundary: convective overshooting (chemical mixing without heat transport), convective penetration (which also transports heat), entrainment of radiative material into the convective zone, and wave- and shear-induced mixing \citep[e.g.][]{saslaw_overshooting_1965,zahn_convective_1991,meakin_turbulent_2007,scott_convective_2021,anders_convective_2023,johnston_modelling_2024}.
In 1D stellar evolution models, this extra mixing beyond the convective core boundary is commonly accounted for by implementing a prescribed overshooting length $\alpha_{\mathrm{ov}}$, the distance at which convective mixing extends beyond the Schwarzschild (or Ledoux) boundary, in units of the pressure scale height \citep[e.g.][]{roxburgh_note_1965,maeder_stellar_1975}.
An alternative parametrization is the exponential diffusive overshooting scheme, which uses a diffusive coefficient $f_{\mathrm{ov}}$, and is motivated by the results of multidimensional simulations that mixing should be gradual rather than a step function.
The latter has grown in popularity, particularly in the asteroseismic community and for the analysis of AGB stars \citep{freytag_hydrodynamical_1996, herwig_evolution_2000,moravveji_tight_2015,moravveji_sub-inertial_2016,buysschaert_forward_2018}.

Many different methods based on empirical information have been used to calibrate these parameters, for example fitting isochrones to star clusters \citep[e.g.][]{maeder_extent_1981,mermilliod_evolution_1986,vandenberg_victoria-regina_2006},
eclipsing binary systems \citep[e.g.][]{schroder_critical_1997,stancliffe_confronting_2015,claret_dependence_2019}, a drop in density of objects in the Hertzsprung--Russell (HR) diagram after the main sequence \citep{castro_spectroscopic_2014,martinet_convective_2021}, a drop in rotational velocities \citep{brott_rotating_2011,martinet_convective_2021},
 asteroseismic observations \citep[for a review, see][]{anders_convective_2023}, the relative number of red and blue supergiants \citep{schootemeijer_constraining_2019}, apsidal motion \citep{rosu_beauty_2026}, nitrogen enrichment \citep{marchant_nitrogen_2026} and various other tests \citep{stothers_observational_1991}.

\begin{figure}[htbp]
	\centering
	\setlength{\abovecaptionskip}{2pt}
	\setlength{\belowcaptionskip}{-4pt}
	\vspace{-6pt}
	\includegraphics[width=\linewidth,trim=13 5 10 10,clip]{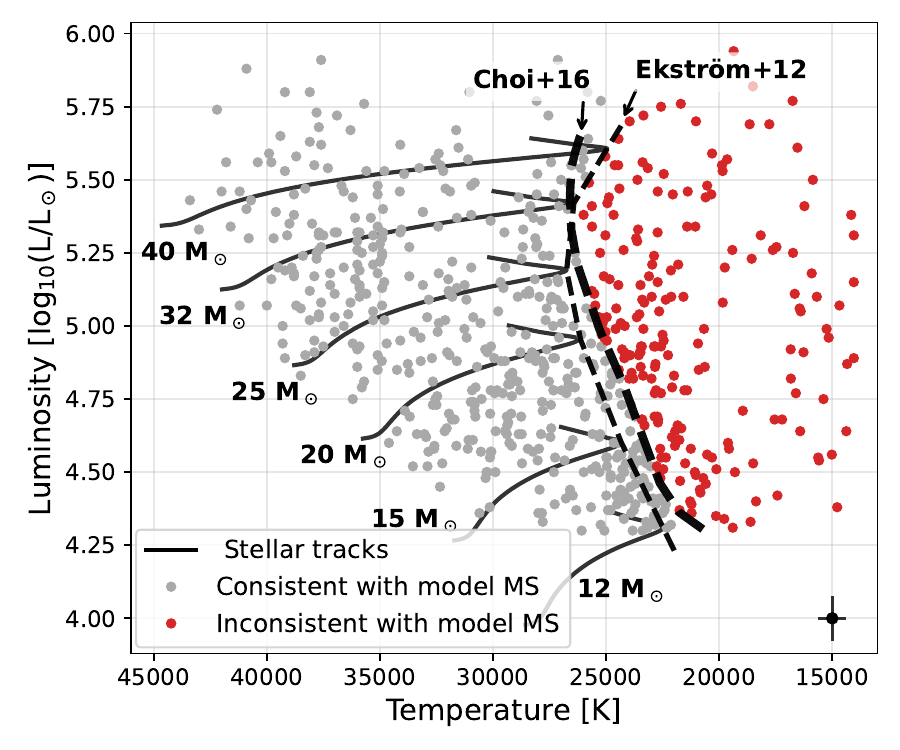}
	\caption{The observed density distribution in the IACOB sample does not match the theoretical end of the main sequence in commonly used stellar evolution models. Model tracks and TAMS boundaries are shown in black solid and dashed lines, respectively. Observed objects are shown in grey and red points. The average observational uncertainty is shown in the bottom right.}
	\label{fig:1}
\end{figure}

Although some of these methods have been applied to stars above $\sim 10$~M$_\odot$, high-mass calibrations remain much scarcer than at lower masses, and the inferred mixing parameters remain highly uncertain.
As a result, prescriptions for overshooting used in large and widely used model sets are often based on calibrations in the low- or intermediate-mass range \citep[e.g.][]{ekstrom_grids_2012,choi_mesa_2016}, which fit modern observations in the high-mass range poorly, as illustrated in Figure \ref{fig:1}.
Such model grids form the basis for widely used applications, including for spectral synthesis codes like STARBURST99 \citep{leitherer_starburst99_1999} and binary population synthesis frameworks such as SEVN \citep{spera_mass_2015,iorio_compact_2023} that can be used to predict gravitational-wave source populations.

On the theoretical side, multidimensional simulations of convection have provided information on the structure of the convectively-unstable core and convective boundary mixing \citep[e.g.][]{kapyla_overshooting_2019,horst_multidimensional_2021,anders_stellar_2022,baraffe_study_2023,andrassy_towards_2024,mao_3d_2024,dethero_shape_2024,kapyla_convective_2024}.
However, translating these results to a treatment suitable for inclusion in 1D stellar evolution models is not straightforward, and the simulations do not include all convective boundary mixing processes, leading to large uncertainties in the predicted overshooting length.
This motivates an observational calibration in the high-mass regime that can directly constrain the effective overshooting length for use in 1D models, agnostic to the relative strengths of the various physical processes occuring at the convective boundary.


Constraining the evolution of massive stars observationally is particularly challenging: they are rare, short-lived, typically distant, and often found in crowded, dust-extinguished star-forming regions \citep{zinnecker_toward_2007,tan_massive_2014}.
For Galactic targets, reliable individual distances were historically difficult to obtain, limiting empirical tests to very small or highly heterogeneous samples.
However, Gaia astrometry now provides robust distances for large numbers of Galactic massive stars \citep[][]{gaia_collaboration_gaia_2016,bailer-jones_estimating_2021,holgado_iacob_2025}, enabling the determination of absolute luminosities and allowing for volume-limited samples \citep[e.g.][]{de_burgos_iacob_2025-1}. At the same time, other observational campaigns are obtaining comparably large samples in the Magellanic Clouds, such as the VFTS or BLOeM surveys \citep{evans_vlt-flames_2011,shenar_binarity_2024}, where distances are well known, but which have other challenges such as crowding or lower signal-to-noise ratios.

The IACOB project has recently presented a homogeneous, statistically meaningful sample of Galactic massive OB stars, mostly consisting of main-sequence objects, making this sample particularly well-suited to study their evolution \citep{simon-diaz_iacob_2011,simon-diaz_iacob_2015,simon-diaz_iacob_2020,simon-diaz_iacob_2026,de_burgos_iacob_2023}.
The transition from slow main-sequence to rapid post-MS evolution produces a drop in HR-diagram number density marking the terminal-age main sequence (TAMS), which \citet{de_burgos_iacob_2025-1} used to propose a TAMS boundary at a roughly constant temperature of $\approx 22$~kK,
which appears to be consistent with a drop in rotation rates \citep[c.f.][]{brott_rotating_2011} and binary fractions \citep[c.f.][]{patrick_binarity_2025}.


This new dataset provides an excellent opportunity to  test and re-calibrate stellar models for massive stars.
Our approach and aims are threefold.

\begin{enumerate}
\item Establish a location of the TAMS using a physics-informed, data-driven method.

\item Determine a prescription for the mixing parameters that leads to a set of models that best reproduces the new data and is consistent with earlier constraints at lower masses.

\item Provide a calibrated set of main-sequence massive star models that can be used for population predictions and the determination of key properties such as the core mass.
\end{enumerate}

To do so, we forward-model the full HR-diagram population and use Bayesian inference, allowing for a wide range of mass-dependent overshooting prescriptions and a background component that accounts for stars not well described by single-star main-sequence evolution.
We then use model comparison to let the data determine which prescription is preferred, in addition to obtaining statistical uncertainties on the overshooting length.
We use the open-source stellar evolution code MESA, in order for our calibrated tracks and overshooting prescription to be easily reproduced, extended, and applied in population and spectral synthesis studies.


\enlargethispage{3\baselineskip}
This paper is organized as follows: in Section \ref{s: Methods}, we describe our methodology for inferring the overshooting length from the observations. In Section \ref{s: Results}, we present our findings. In Section \ref{s: Discussion}, we compare with other work and discuss the implications of our results. Section \ref{s: Conclusion} summarizes our conclusions.

\section{Method}\label{s: Methods}
\subsection{Data}\label{ss: Data}
We use the most complete version of the IACOB database (\citealt{simon-diaz_iacob_2020}, see also \citealt{holgado_iacob_2020,de_burgos_iacob_2023}), which comprises a homogeneous selection of Galactic massive stars, including roughly 900 single stars and single-lined spectroscopic binary systems within 4~kpc, described in detail in \citet{de_burgos_iacob_2025-1}.
For the purposes of this study, we use a subset of the sample, consisting of stars within 2.5~kpc, which is a good trade-off between completeness and sample size \citep[see][]{de_burgos_iacob_2025-1}.
The sample includes likely single stars and single-lined spectroscopic binary systems, and exclude hypergiants, Be stars, and other disk-bearing objects, because their spectra are challenging or not reliably modeled with current 1D atmospheric analyses. Double- or higher-order spectroscopic binaries are also excluded because these systems have not yet been fully analyzed in the IACOB framework \citep{de_burgos_iacob_2024}.
We restrict the sample to $\log(L/L_\odot) > 4.5$. Below this luminosity, the IACOB sample contains
much fewer slow rotators, which, as reported in \citet{de_burgos_iacob_2025-1}, could by caused by selection effects at low $L$. Therefore, we exclude that regime, leaving 604 stars in our analysis.



\subsection{MESA models}\label{ss: Models}
In order to model a synthetic population of stars, we use the stellar evolution code MESA version \texttt{r23.05.1} \citep{paxton_modules_2011,paxton_modules_2013,paxton_modules_2015,paxton_modules_2018,paxton_modules_2019,jermyn_modules_2023}.
Our models adopt a protosolar composition with Z = 0.0154 \citep{asplund_chemical_2021}, the `Cox' mixing-length theory prescription \citep{cox_principles_1968} using the Ledoux criterion \citep{langer_semiconvective_1983,choi_mesa_2016}, and a 23-isotope nuclear network suitable to capture the main-sequence evolution. We use the empirical wind mass-loss prescription of \citet{pauli_new_2025}, where the mass-loss rate is a function of the electron-scattering Eddington parameter.
Our rotating models, which have an initial equatorial rotation rate of 200~km~s$^{-1}$, additionally include rotational mixing and angular-momentum transport \citep{heger_presupernova_2000,heger_presupernova_2005,jin_boron_2024}.
Further configuration details are given in Appendix \ref{sec:appendix_mesa_config}.

We compute the main-sequence evolution of the stars using a grid of 800 models, varying initial mass from 12 to 60 M$_\odot$ taking 40 steps in logarithmic mass, and varying the mixing parameters from 0 to 0.5 or 0 to 0.05 for step and exponential overshooting respectively in 20 linear steps.
We use linear interpolation between the grid points using mass and fractional main-sequence lifetime.

Additionally, we explore different variations: using two alternative wind prescriptions, considering rotation, and changing the treatment of convection in radiation-dominated convective layers (known as MLT++), which prevents the development of locally super-Eddington radiative luminosities (\citealt{paxton_modules_2013}, see also \citealt{sanyal_massive_2015}).
\begin{figure*}[htbp]
	\centering
	\includegraphics[width=\textwidth]{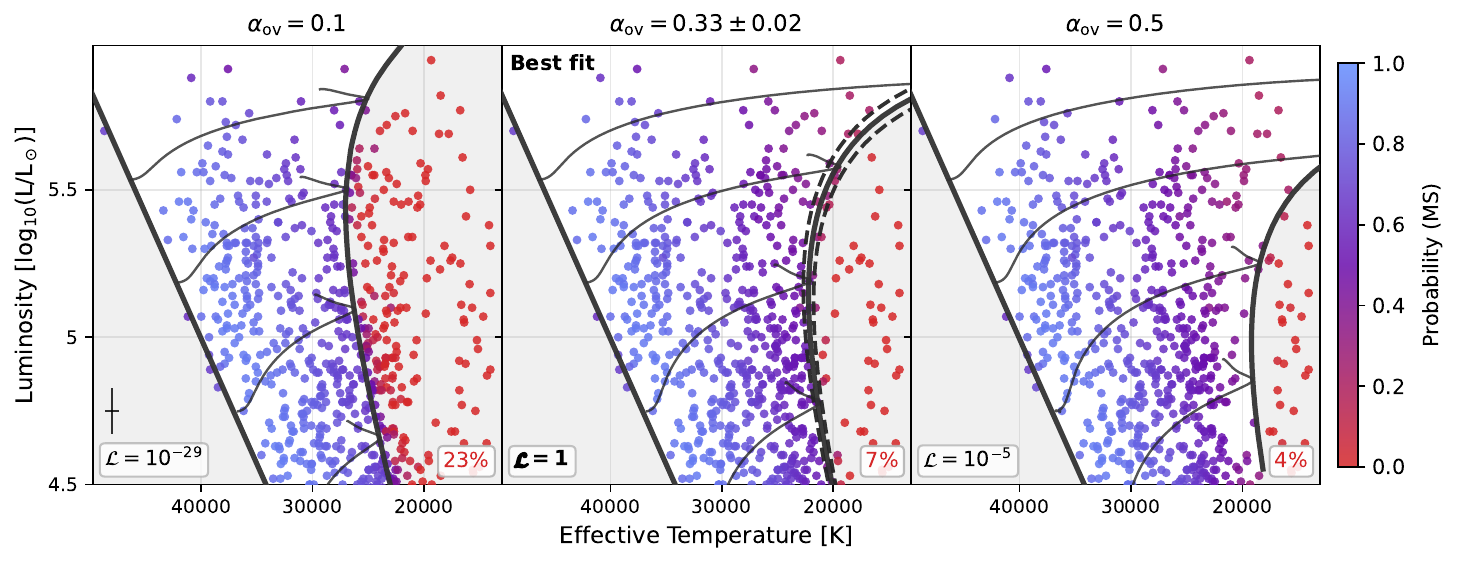}
	\caption{Varying the overshooting length and comparing a synthetic population of stars to the observed data in the HR diagram finds a best-fit overshooting length of $\alpha_{\rm ov} = 0.33 \pm 0.02$. The likelihood of the different overshooting values is shown in the bottom left of each panel, where the middle panel shows the highest likelihood. The color of the observed stars represents the probability of being part of the main-sequence population.
	The percentage of observed stars that have a probability lower than 5\% is shown in the bottom right of each panel. Four main-sequence tracks are shown in light grey for reference, with masses of 15, 22, 34, and 49 M$_\odot$.
	}
	\label{fig:2}
\end{figure*}

\subsection{Inference}\label{ss: Inference}
We model the distribution of stars in the HR diagram as a mixture of a main-sequence component and a background component, which may include certain products of binary mass transfer or merger remnants.
We allow for a mass-dependence of the convective boundary mixing parameters, including linear, exponential, and polynomial dependencies, with the goal of letting the data decide which dependence is most appropriate.

We use a mixture likelihood approach to model both a regular main-sequence component ($\mathcal{L}_{\rm MS}$), which is described by single stellar physics, and a possible background component ($\mathcal{L}_{\rm bg}$), which allows for various contaminants of stars that are not well described by our main-sequence component.
The fitted mass-dependent mixing parameters in the inference are  $\theta$ = $\alpha_{\rm ov}(M)$ or $f_{\rm ov}(M)$ for step and exponential overshooting respectively.
The observed data for each star $i$ in the HR diagram are represented by the vector $\mathbf{y}_i = (T_{{\rm eff},i}, \log L_i)$.
The likelihood of that star being described by the main-sequence component is given by:
\begin{equation}
	\mathcal{L}_{{\rm MS},i}
	=	\int p\!\left(\mathbf{y}_i \mid M,t,\theta\right)\,
		 p\!\left(M,t \mid \theta\right)\,
	\, dM\, dt ,
\end{equation}

\noindent where $t$ and $M$ are the age and mass of the star, and where the observational likelihood $p\!\left(\mathbf{y}_i \mid M,t,\theta\right)$ is a 2D Gaussian around the observed data point $\mathbf{y}_i$, with standard deviation given by the observational uncertainties.
 $p\!\left(M,t \mid \theta\right)$ is the population prior along the main sequence, for which we assume a Salpeter IMF \citep{salpeter_luminosity_1955} and a continuous star formation history. We apply a cut of $\log(L/L_\odot) > 4.5$ to the population prior, consistent with the luminosity range of our sample.


For the background component $\mathcal L_{{\rm bg}}$, we assume a uniform distribution in $(\log T_{\rm eff},\log L)$, with A$_{\rm HR}$ the rectangular area in the HR diagram spanned by the observed sample, leading to
$
\mathcal L_{{\rm bg}}
=1/{A_{\rm HR}}.$
The full-sample likelihood is then given by:
\begin{equation}
	\mathcal{L}(\theta,\pi_{\rm bg})
	=
	\prod_{i=1}^{N}
	\left[
	(1-\pi_{\rm bg})\,\mathcal{L}_{{\rm MS},i}(\theta)
	+
	\pi_{\rm bg}\,\mathcal{L}_{{\rm bg}}
	\right],
\end{equation}

\noindent where $\pi_{\rm bg}$ is a relative weighting term $\in [0,1]$ for the background component and is a free parameter in the inference, and N is the total number of stars in the sample.
Additionally, although there are stars with apparent masses higher than 40 M$_\odot$, we do not include them in our main analysis, as a moderate step overshooting length of 0.35 leads to the main sequence ending at temperatures below 15,000 K, which is close to the temperature limit of the observational data. In addition, the sample gets increasingly scarce at higher masses, becoming more prone to unwanted stochastic effects.
Instead, we let stars with masses between 40 and 60 M$_\odot$ have their own constant overshooting length, as a free parameter, in order to not include them as part of the `background component'.

Inference is performed with the \texttt{ultranest} nested sampling package \citep{buchner_ultranest_2021}, which implements the MLFriends algorithm \citep{buchner_statistical_2016,buchner_collaborative_2019} to sample the posterior distribution and Bayesian evidence.
Uniform priors are adopted for all parameters within their physical range, with overshooting length $\alpha_{\rm ov} \sim U[0,0.5]$ and background weighting term $\pi_{\rm bg} \sim U[0,1]$.

\section{Results}\label{s: Results}
Figure \ref{fig:2} illustrates how the end of the main sequence shifts as a function of the overshooting length when adopting a constant value, here shown for a step overshooting scheme.
The best-fit overshooting length that maximizes the likelihood of the observed data is $\alpha_{\rm ov} = 0.33 \pm 0.02$ and is shown in the middle panel. Both the left and right panels are disfavored, as they fit the observed data poorly, but are shown for reference.

\subsection{Mass dependence}
\begin{figure*}[htbp]
	\centering
	\includegraphics[width=0.85\textwidth]{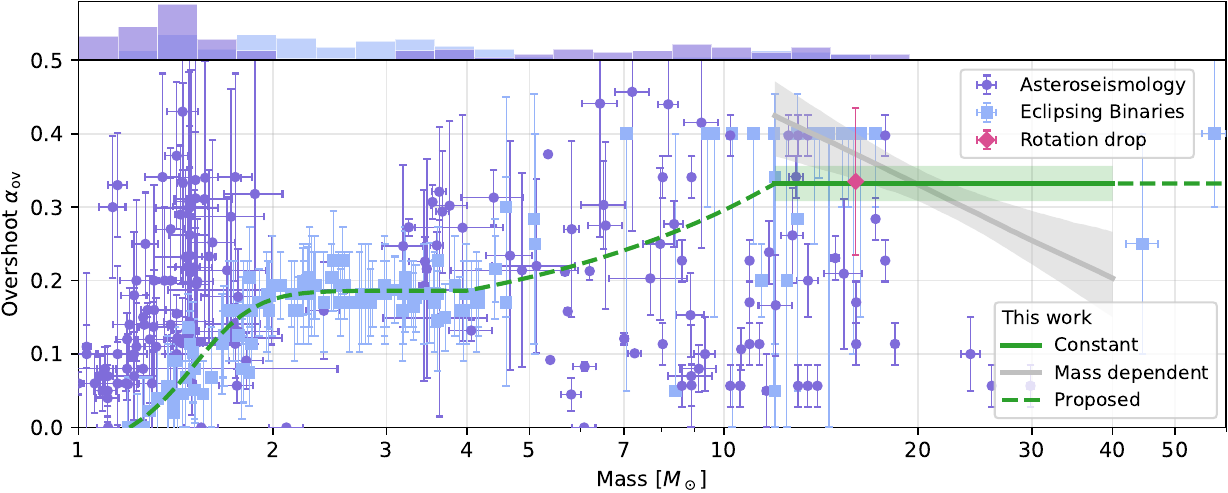}
	\caption{A comparison of a wide range of observationally derived constraints on overshooting (from asteroseismology, eclipsing binaries, and rotational velocities), compared to our best-fit constant value, and our best-fit mass-dependent relation. The latter is not strongly preferred over the constant value. Our proposed prescription from Equation \ref{eq:alpha_ov_prescription} for the whole mass range is also shown in green dashed lines.}
	\label{fig:3}
\end{figure*}

Low- and intermediate-mass stars show a clear increase in overshooting with mass (e.g. \citealt{claret_dependence_2019}; see also Figure~\ref{fig:3}),
as do predictions from simulations \citep{baraffe_study_2023,johnston_modelling_2024}, and some previous studies at high mass \citep[e.g.][]{castro_spectroscopic_2014}.
Therefore, we extend our analysis beyond a constant prescription and test linear, log-linear, exponential, polynomial, and piecewise-constant dependencies with 2--5 mass segments.
In all these experiments, we find that the best-fitting mass-dependence favors a decrease of overshooting with increasing mass.
Strikingly, however, Bayesian model comparison shows that constant and decreasing prescriptions receive comparable support in the 12--40~M$_\odot$ range, once model complexity is penalized through the evidence and Bayes factors, with neither prescription strongly preferred over the other.
For simplicity, we adopt a constant overshooting length for the main analysis. The full set of mass-dependent experiments is described in Appendix~\ref{sec:appendix_mass_dependent_overshooting}.

\begin{figure}[htbp]
	\centering
		{\includegraphics[width=1.0\columnwidth]{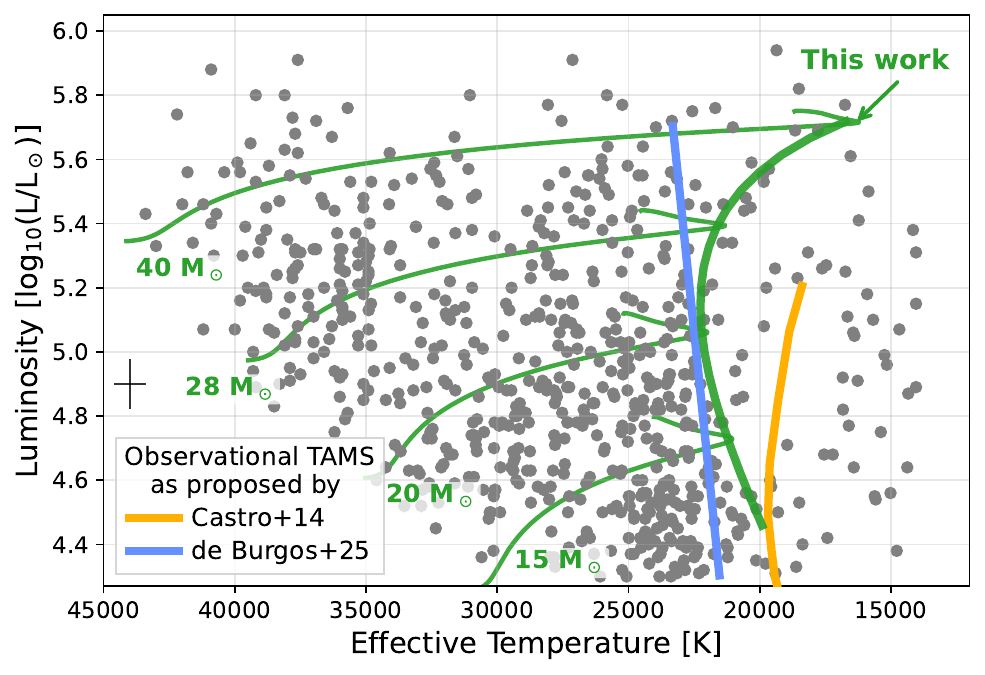}}
	\caption{Comparison of the location of our proposed TAMS in the HR diagram, with prior work.
	We include the TAMS proposed by \citet{de_burgos_iacob_2025-1} for this dataset, as well as the one proposed by \citet{castro_spectroscopic_2014} based on spectroscopic data only. }
	\label{fig:obsTAMS}
\end{figure}

\subsection{Robustness of our findings}
When examining the sensitivity of our results to changes in the physical assumptions of the stellar evolution models, we found that our best-fit overshooting length is robust.
To test this, we recomputed dedicated MESA grids that each change one ingredient relative to the default setup, and repeated the full inference for each grid.
In particular, when replacing the step overshooting scheme with an exponential one, we obtained $f_{\rm ov} = 0.028 \pm 0.003$.
Converting this to an equivalent step-overshooting length with the $\approx 1{:}11$ factor of \citet{claret_dependence_2017} yields a value that is consistent with our default $\alpha_{\rm ov}$.
Neither changing the empirical wind mass-loss rates from \citet{pauli_new_2025} to the theoretical models from \citet{bjorklund_new_2023}, nor enabling MLT++ \citep{paxton_modules_2013} changed the best-fit overshooting length outside of the uncertainty range (see further details in Appendix \ref{sec:appendix_mass_dependent_overshooting}).
When using rotating models with an initial equatorial rotation rate of 200~km~s$^{-1}$, we found an increased uncertainty on the lower bound of the overshooting length, where $\alpha_{\rm ov} = 0.334^{+0.008}_{-0.045}$.
This is likely due to the fact that rotational mixing contributes to convective boundary mixing, which can mimic the effect of an increased core overshooting.
We note that the sample is mostly composed of slow rotators ($v\sin i < 100$~km~s$^{-1}$), so rapidly rotating models might not provide a representative description of the bulk of the population.
When using the Dutch wind scheme, which combines the mass-loss rates from \citet{vink_mass-loss_2001} and \citet{de_jager_mass_1988} in the prescription of \citet{glebbeek_evolution_2009}, we found a lower overshooting length of $\alpha_{\rm ov} = 0.228 \pm 0.013$. For this wind scheme, the end of the main sequence is shifted to lower temperatures, which leads to a lower inferred overshooting length to compensate for this effect.
Finally, we found that when changing the distance threshold of the observed sample from 2.5~kpc to 2~kpc (aiming at increasing the completeness within the Galactic volume; see \citealt{de_burgos_iacob_2025-1}), and thereby reducing the sample size by roughly one third, the best-fit overshooting length remains consistent, within uncertainties. 

\subsection{Location of the Terminal Age Main Sequence}
A key output of our analysis is a physically motivated, data-driven location of the TAMS in the HR diagram (Figure~\ref{fig:obsTAMS}).
We compare our results with two earlier determinations.
Before Gaia distances were available, \citet{castro_spectroscopic_2014} placed the TAMS using spectroscopic data and the associated spectroscopic HR diagram \citep{langer_spectroscopic_2014}.
More recently, \citet{de_burgos_iacob_2025-1} used a  sample similar to ours to locate the TAMS. To do so, they used four luminosity bins and a 15\% threshold in the cumulative effective temperature distribution to find a drop in the number of stars after the TAMS, which appears to be at a roughly constant temperature of $\approx 22$~kK.
Around 20~M$_\odot$, our inferred TAMS agrees well with \citet{de_burgos_iacob_2025-1}, deviating at lower and higher masses.
Mapping the \citet{castro_spectroscopic_2014} HR diagram to a bolometric scale with our best-fit models, we find our TAMS shifted to higher temperatures, especially at high masses, highlighting a major revision relative to this older calibration.



\subsection{Our mass-dependent overshooting prescription}\label{ss:prescription}
In Figure~\ref{fig:3}, we compare our best-fit overshooting length (from both a constant and a mass-dependent inference run) with independent observational constraints from asteroseismology \citep{moravveji_sub-inertial_2016,noll_probing_2021,anders_convective_2023,vanrespaille_asteroseismic_2026}, eclipsing binaries \citep{claret_dependence_2016,claret_dependence_2017,claret_dependence_2018,claret_dependence_2019,higgins_massive_2019,johnston_modelling_2019,tkachenko_mass_2020}, and an observed drop in rotational velocities \citep{brott_rotating_2011}.
At the high-mass end, constraints are scarce, which is the region covered by our analysis.
Although individual determinations show considerable scatter, especially in asteroseismology \citep{aerts_forward_2018,johnston_modelling_2019,johnston_one_2021}, the ensemble suggests a general increase in overshooting length with mass, from $\approx 0.1$--$0.2$ at 2~M$_\odot$ to $\approx 0.3$ at 20~M$_\odot$. 

Combining these lower-mass calibrations with our high-mass results, we propose a prescription from 1.2 to 40~M$_\odot$.
For step overshooting, $\alpha_{\rm ov}(M)$ is given by
\begin{equation}\label{eq:alpha_ov_prescription}
\alpha_{\rm ov}(M)=
\begin{cases}
11.36\!\left(\dfrac{0.02013}{1+e^{-5.5(M-1.47)}}-0.00373\right),\\
\qquad \qquad \qquad  \text{if} \quad 1.2 M_\odot \le M \le 4M_\odot;\\[0.6ex]
0.113 + 0.0183\,M,\\  
\qquad \qquad \qquad  \text{if} \quad 4 M_\odot < M < 12M_\odot;\\[0.6ex]
0.332,\\
\qquad \qquad \qquad  \text{if} \quad 12 M_\odot \le M,
\end{cases}
\end{equation}
\noindent where $M$ is in units of~$M_\odot$, the low-mass branch follows the empirical eclipsing-binary relation from \citet{claret_dependence_2018}, the high-mass branch follows our empirical best-fit value, and the intermediate-mass branch is a linear interpolation between the two.
For an exponential diffusive overshooting parametrization, we find
\begin{equation}\label{eq:fov_prescription}
	f_{\rm ov}(M)=
	\begin{cases}
	\dfrac{0.02013}{1+e^{-5.5(M-1.47)}}-0.00373,\\
	\qquad \qquad \qquad  \text{if} \quad 1.2 M_\odot \le M \le 4M_\odot;\\[0.6ex]
	0.0107 + 0.00144\,M,\\  
	\qquad \qquad \qquad  \text{if} \quad 4 M_\odot < M < 12M_\odot;\\[0.6ex]
	0.0279,\\
	\qquad \qquad \qquad  \text{if} \quad 12 M_\odot \le M.
	\end{cases}
	\end{equation}

\begin{figure*}[htbp]
	\centering
	\subcaptionbox{Comparison with theoretical predictions from convection simulations and models.\label{fig:4b}}%
		{\includegraphics[width=0.85\textwidth]{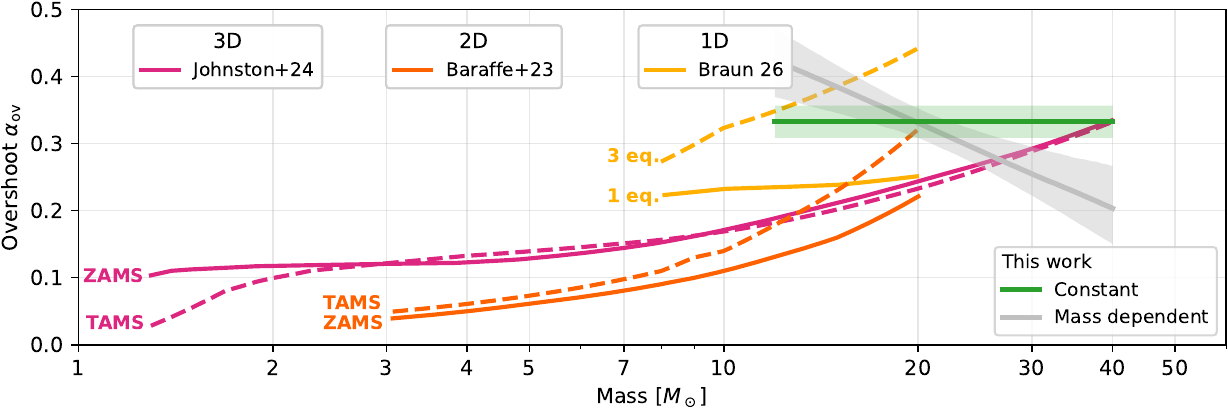}}\\[0.6ex]
	\subcaptionbox{Comparison with commonly used stellar evolution code prescriptions.\label{fig:4c}}%
		{\includegraphics[width=0.85\textwidth]{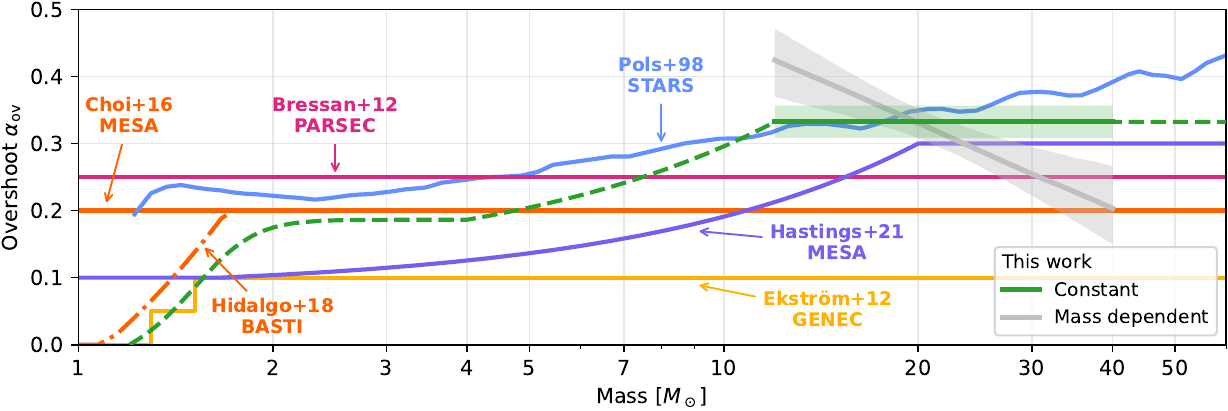}}
	\caption{Comparison of our best-fit overshooting value with (a) predictions from convection simulations and (b) other common prescriptions used in stellar evolution models shown in solid lines, including our own proposed prescription in dashed lines.}
	\label{fig:4}
\end{figure*}

\section{Discussion}\label{s: Discussion}
\subsection{Comparison with theoretical results}
In Figure \ref{fig:4}, we show a comparison to predictions from convection models, and other commonly used stellar evolution code prescriptions.
Panel (a) shows three different theoretical predictions from convection simulations and models, which generally show an increasing overshooting length with mass \citep{baraffe_study_2023,johnston_modelling_2024,braun_application_2026}. 
The models range from 1D to 3D, all using different physical assumptions and numerical methods, meaning they are not directly comparable.
For example, \citet{johnston_modelling_2024}, which is based on \citet{anders_stellar_2022} (see also \citealt{jermyn_convective_2022}), isolate the effect of convective penetration and \citet{baraffe_study_2023} does not include 3D geometries.
Both studies consider non-rotating models, but rotating stars are expected to have a higher overshooting length due to rotationally induced shear mixing.
Additionally, tidal interactions in close binaries could enhance or suppress mixing, depending on the angular momentum transport assumptions \citep{hastings_internal_2020,koenigsberger_induced_2021,sciarini_chemical_2026}.
In a 1D model, \citet{braun_application_2026} used the turbulent convection model proposed by \citet{kuhfus_modell_1987} \citep[see also][]{kupka_stellar_2022,ahlborn_stellar_2022,braun_testing_2024,braun_testing_2026}, aimed at improving convective modeling in 1D stellar evolution codes. It comprises a 1-equation and a more physically complete 3-equation model.

In general, but particularly below 20 M$_\odot$, our best-fit overshooting length is consistently higher than the predictions from the simulations. This is likely due to the fact that the simulations do not intend to capture all convective boundary mixing mechanisms, while observationally calibrated values serve as a proxy for their combined effect.

\subsection{Comparison with other overshooting prescriptions}
Panel (b) of Figure \ref{fig:4} shows commonly used prescriptions in stellar evolutionary codes, including ones used in MESA \citep{choi_mesa_2016,hastings_stringent_2021}, STARS \citep{pols_stellar_1998}, GENEC \citep{ekstrom_grids_2012}, PARSEC \citep{bressan_parsec_2012}, and BASTI \citep{hidalgo_updated_2018}. 
Some have historically been calibrated on low-mass stars, such as the GENEC prescription which is used in spectral synthesis models like STARBURST99 \citep{leitherer_starburst99_1999,leitherer_effects_2014}, but this does not fit well for high-mass stars.
Others include a mass dependence, to account for the predicted and observed mass dependence of the overshooting length, such as the ones in BASTI and the MESA prescription from \citet{hastings_stringent_2021}, which is used in the detailed binary models by \citet{jin_comprehensive_2026}. \citet{pols_stellar_1998} do not parametrize the overshooting length, and instead base their treatment on the stability criterion itself. Here, mixing occurs in regions which would be classically stable, but only if they are stable by a small margin. This naturally leads to an increasing overshooting length with mass and is used in BPASS \citep{eldridge_binary_2017,byrne_bpass_2025}. This prescription, in addition to the one from \citet{hastings_stringent_2021}, is closest to our best-fit overshooting value.

\subsection{Comparison of the TAMS location with other models}
\begin{figure}[htbp]
	\centering
	\includegraphics[width=1.0\columnwidth]{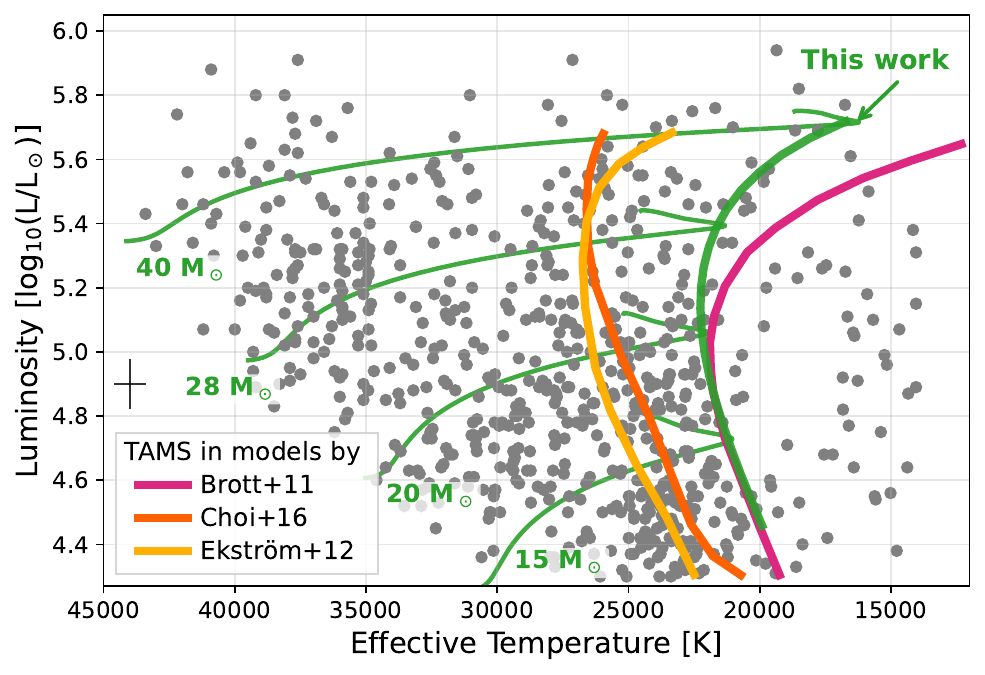}
	\caption{Comparison of the location of our physically-motivated, data-driven TAMS in the HR diagram, with other common stellar models.}
	\label{fig:5}
\end{figure}

In Figure \ref{fig:5}, we compare the location of the TAMS in the HR diagram between various commonly used stellar tracks \citep{brott_rotating_2011,choi_mesa_2016,ekstrom_grids_2012} and our work.
Compared to the tracks from \citet{ekstrom_grids_2012} and \citet{choi_mesa_2016}, our TAMS is shifted to lower temperatures.
Note that our overshooting length is almost identical to the one from \citet{brott_rotating_2011}, but an important difference comes from the updated wind scheme from \citet{pauli_new_2025}.




\subsection{Comparison of the helium core mass at the TAMS with other models}
\begin{figure}[htbp]
	\centering
	\includegraphics[width=1.0\columnwidth]{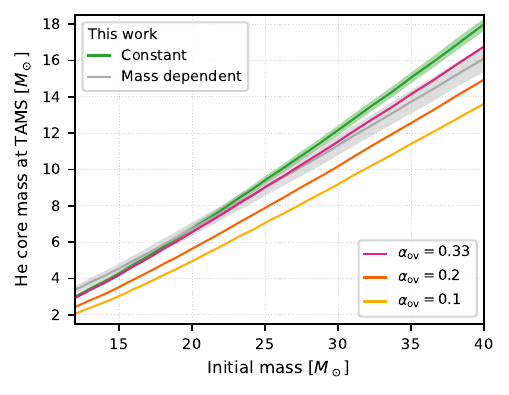}
	\caption{Comparison of the helium core mass at the TAMS of our best-fit models (using the stellar winds from \citet{pauli_new_2025}, in green and grey lines, including uncertainties), compared to models with different overshooting values (using the winds from \citet{vink_mass-loss_2001}, in yellow, orange and purple lines).}
	\label{fig:6}
\end{figure}

In Figure~\ref{fig:6}, we compare the helium-core mass at the TAMS in our best-fit models with models that adopt overshooting lengths and wind prescriptions matching those of \citet{brott_rotating_2011}, \citet{ekstrom_grids_2012}, and \citet{choi_mesa_2016}.
We define the helium-core mass as the mass interior to the point where the hydrogen mass fraction drops below 0.01 and the helium mass fraction exceeds 0.1.
We find that our best-fit model with a constant overshooting length of $\alpha_{\rm ov} = 0.33$ has the highest helium-core mass of the models shown.
Relative to those models, our helium-core masses are larger by a few percent up to 40\%.
This is in part due to the higher overshooting length and an updated, lower wind mass-loss rate for massive stars (\citealt{pauli_new_2025} rather than \citealt{vink_mass-loss_2001}).

\subsection{Caveats}
An important caveat of our inference is the adopted contamination model. We describe contaminants with a uniform background, which intends to capture the broad range of possible contaminants such as post-binary-interaction products, evolved stars close to the TAMS, or survey-selection features that cluster in specific regions of the HR diagram. As these are likely not uniformly distributed, this may lead to a bias in the inferred overshooting length.
A useful extension of this work is therefore to test more flexible background models (e.g., mixture distributions using empirically or simulation informed templates) and propagate the resulting model uncertainty into the overshooting calibration. In our experiments, we infer values for the mixture weighting $\pi_{\rm bg}$ of approximately 30\%, which indicates that the
population fit needs a considerable component from stars which are not well-described by the evolution of isolated main-sequence stars.

Additionally, we assumed that overshooting is constant throughout the main-sequence lifetime.  However, multi-dimensional simulations find that as the core grows, the overshooting length changes \citep{baraffe_study_2023,johnston_modelling_2024}. We are therefore modeling the effective overshooting length over the main-sequence lifetime.
We also assumed that the same overshooting length applies to stars with the same mass \citep[c.f.][]{johnston_one_2021}. However, depending on factors like the composition, binarity, rotation rate, or magnetic field strength of the star, the amount of convective boundary mixing changes. Studying the effect of these factors in further detail than our current analysis, would be an interesting extension of our work.

\section{Conclusions}\label{s: Conclusion}

There is a pressing need in the community for well-calibrated models of massive stars.
Many widely used evolutionary grids still rely on convective boundary mixing prescriptions extrapolated from lower-mass calibrations, and as a result they reproduce the observed distribution of stars in the HR diagram poorly (Figure~\ref{fig:1}).
At the same time, Gaia astrometry and the homogeneously analyzed IACOB spectroscopic sample of Galactic OB stars now make it possible to forward-model the HR-diagram population of massive stars at a scale that is unprecedently large.
Using this dataset together with MESA models and Bayesian inference, we draw the following conclusions:

\begin{itemize}
\item In the mass range 12--40~M$_\odot$, a constant convective-boundary mixing parameter provides the best match to the data, with $\alpha_{\mathrm{ov}} = 0.33 \pm 0.02$ for step overshooting or $f_{\mathrm{ov}} = 0.028 \pm 0.003$ for exponential overshooting. This calibration is robust to moderate changes in the adopted stellar-physics assumptions and sample selection.

\item The data disfavor a continuing rise in overshooting with mass in this regime, but allow a decrease with mass. Bayesian model comparison shows comparable support for constant and decreasing prescriptions, in contrast to the clear increase found at lower masses.

\item Our analysis yields a physically motivated, data-driven location of the TAMS in the HR diagram (Figure~\ref{fig:obsTAMS}).

\item Combining our high-mass constraints with earlier calibrations, we propose a mass-dependent overshooting recipe from 1.2 to 40~M$_\odot$ (Equations~\ref{eq:alpha_ov_prescription} and \ref{eq:fov_prescription}).

\item Compared to several commonly used prescriptions, including those adopted in STARBURST99, our calibration implies systematically larger helium-core masses (Figure~\ref{fig:6}), with corresponding shifts in the location of the TAMS (Figure~\ref{fig:5}).

\item We provide a set of pre-computed, calibrated main-sequence MESA models that can be used for a wide range of applications, available on Zenodo\footnote{\url{https://doi.org/10.5281/zenodo.21496755}}.
\end{itemize}

These results have broad implications for applications that rely on  stellar model grids.
Spectral synthesis and population synthesis codes that adopt lower overshooting prescriptions than we infer will shift predicted HR-diagram densities and bias inferred rates and properties of stellar end products, including compact remnants and gravitational-wave sources.
Because core growth during the main sequence affects the mapping between initial and core mass, it will also affect predictions for the red-supergiant fraction and for binary channels that depend on the timing of mass transfer.
Future work should test whether the same calibration holds at lower metallicity by extending this analysis to similarly complete samples in the LMC and SMC.

\section*{Acknowledgements}
We would like to thank Mathieu Renzo, Ylva Götberg, Norbert Langer, Abel Schootemeijer, Gijs Nelemans, Teresa Braun, and Jim Fuller for useful discussions.
This research was supported in part by grant NSF PHY-2309135 to the Kavli Institute for Theoretical Physics (KITP).
This research has made use of the Astrophysics Data System, funded by NASA under Cooperative Agreement 80NSSC21M00561.
S.S-D acknowledges the support from the State Research Agency (AEI) Spanish Ministry of Science and Innovation (MICIN) and the European Regional Development Fund (FEDER) under grant Productos de la interacción de estrellas masivas revelados por grandes sondeos espectroscópicos, with reference PID2024-159329NB-C21. S.S-D also acknowledges funding from European Commission (EC) under Project OCEANS - Overcoming challenges in the evolution and nature of massive stars, HORIZON-MSCA-2023-SE-01, No G.A 101183150 Funded by the European Union.

\section*{Software and Data}
This work made use of the following software packages: \texttt{python} \citep{van_rossum_python_2009}, \texttt{numpy} \citep{harris_array_2020}, \texttt{matplotlib} \citep{hunter_matplotlib_2007}, \texttt{scipy} \citep{virtanen_scipy_2020, gommers_scipyscipy_2025}, and Modules for Experiments in Stellar Astrophysics (\texttt{MESA}) \citep{paxton_modules_2011, paxton_modules_2013, paxton_modules_2015, paxton_modules_2018, paxton_modules_2019, jermyn_modules_2023}.
The complete MESA inlist files and stellar tracks are available on Zenodo at \url{https://doi.org/10.5281/zenodo.21496755}.

\bibliography{bibliography}{}
\bibliographystyle{aa}

\begin{appendix}
\nolinenumbers





\section{Stellar evolution models}
\label{sec:appendix_mesa_config}
To model the main-sequence evolution of the massive stars in our sample, we employ the  one-dimensional stellar evolution code MESA \citep{paxton_modules_2011, paxton_modules_2013, paxton_modules_2015, paxton_modules_2018, paxton_modules_2019, jermyn_modules_2023} version \texttt{r23.05.1}.
The key physical assumptions of our models are summarized below:

\begin{itemize}
\item We adopted a protosolar metallicity of Z = 0.0154 with an initial helium mass fraction of Y = 0.2725 \citep{asplund_chemical_2021}.

\item For convective boundaries, we used the  `Cox' mixing length theory \citep{cox_principles_1968} with $\alpha_\mathrm{MLT}$ = 1.82 \citep{choi_mesa_2016} (see also \citealt{joyce_review_2023}) and the Ledoux criterion for convective stability with a semiconvection efficiency parameter of $\alpha_\mathrm{sc}$ = 1.0 \citep{langer_semiconvective_1983}.

\item We used a nuclear reaction network with 23 isotopes (`approx21\_cr60\_plus\_co56.net'), which includes the main reactions relevant for hydrogen and helium burning, as well as for later stages if needed.

\item In our default models, mass loss is included via the empirical wind prescription of \citet{pauli_new_2025}, which is calibrated over the mass and temperature range of our models and depends on the electron-scattering Eddington parameter $\Gamma_{\rm e}$.
Compared with commonly used temperature-based recipes for OB stars \citep[such as][]{vink_mass-loss_2001}, this scheme yields mass-loss rates that are lower by roughly an order of magnitude and does not produce a sharp increase in mass loss at a particular temperature.
We based the stellar-wind portions of our MESA inlists on those of \citet{romagnolo_stellar_2026}.

\item In our rotating models, we include rotational mixing with an inhibiting factor $f_\mu = 0.05$  \citep{heger_presupernova_2000} and a diffusion coefficient $f_c = 0.017$ \citep{jin_boron_2024}. We also include Eddington-Sweet circulation, secular shear instability, Solberg-Høiland instability, and Goldreich-Schubert-Fricke instability \citep{heger_presupernova_2000,heger_presupernova_2005}.

\end{itemize}

For each of our setups, we compute the evolution of 800 single-star MESA models with Zero-age Main Sequence (ZAMS) masses between 12 and 60 M$_\odot$ with 40 log-spaced increments; step overshooting lengths from 0 to 0.5 and exponential overshooting lengths from 0 to 0.05 in 20 linearly spaced increments. We run the models until the end of core helium burning, or until the model reaches 50000 steps.
We then focus on the main-sequence phase of the models, and use linear interpolation to derive a mapping from initial mass, overshooting length, and (fractional main sequence) age, to the position of the model in the HR diagram. This mapping is used to create the synthetic population of stars, which is then compared to the observed data in the main text.
Our synthetic population is created by sampling a Salpeter initial mass function with a slope of -2.35 \citep{salpeter_luminosity_1955}, assuming continuous star formation, and a luminosity cut of $\log(L/L_\odot) > 4.5$.

\section{Mass-dependent overshooting experiments}
\label{sec:appendix_mass_dependent_overshooting}

We compare various mass-dependent overshooting prescriptions using Bayesian inference, for six stellar-evolution grids and two distance cuts. The following subsections provide more details on the prescriptions, grids, and the inference setup.

\subsection{Overshooting prescriptions}\label{ss:appendix_overshoot_prescriptions}
An overview of the mass-dependent overshooting prescriptions can be found in Table~\ref{tab:overshoot_prescriptions}, where their names, degrees of freedom, and analytical forms are listed.
Namely, within the mass range of 12 to $40M_\odot$, we tested a constant overshooting value, a linear dependence, a linear dependence in log mass, a quadratic dependence, an exponential dependence, and piecewise constant values in $n$-log spaced mass bins, with $n$ ranging from 2 to 5.

\begin{table*}[htbp]
	\centering
	\begin{tabular}{@{}l@{\quad}c@{\quad}c@{}}
	\hline\hline
	Prescription & Degrees of freedom & $\alpha_{\rm ov}(M)=$ \\
	\hline
	constant & 1 & $\alpha$ \\
	linear & 2 & $\alpha + \beta M$ \\
	log $M$ linear & 2 & $\alpha + \beta \log M$ \\
	exponential & 3 & $\alpha + \beta \exp(\gamma M)$ \\
	quadratic & 3 & $\alpha + \beta_1 M + \beta_2 M^2$ \\
	piecewise constant, $n=2$ & 2 & $\displaystyle
	\begin{cases}
	\alpha_1 & M < 21.9\,M_\odot \\
	\alpha_2 & M \ge 21.9\,M_\odot
	\end{cases}$ \\
	piecewise constant, $n=3$ & 3 & $\displaystyle
	\begin{cases}
	\alpha_1 & M < 17.9\,M_\odot \\
	\alpha_2 & 17.9\,M_\odot \le M < 26.8\,M_\odot \\
	\alpha_3 & M \ge 26.8\,M_\odot
	\end{cases}$ \\
	piecewise constant, $n=4$ & 4 & $\displaystyle
	\begin{cases}
	\alpha_1 & M < 16.2\,M_\odot \\
	\alpha_2 & 16.2\,M_\odot \le M < 21.9\,M_\odot \\
	\alpha_3 & 21.9\,M_\odot \le M < 29.6\,M_\odot \\
	\alpha_4 & M \ge 29.6\,M_\odot
	\end{cases}$ \\
	piecewise constant, $n=5$ & 5 & $\displaystyle
	\begin{cases}
	\alpha_1 & M < 15.3\,M_\odot \\
	\alpha_2 & 15.3\,M_\odot \le M < 19.4\,M_\odot \\
	\alpha_3 & 19.4\,M_\odot \le M < 24.7\,M_\odot \\
	\alpha_4 & 24.7\,M_\odot \le M < 31.4\,M_\odot \\
	\alpha_5 & M \ge 31.4\,M_\odot
	\end{cases}$ \\
	\hline
	\end{tabular}
	\caption{Mass-dependent overshooting prescriptions used in the inference experiments. Knot masses for piecewise-constant prescriptions are fixed at $\log_{10}$-spaced edges from 12 to 40~$M_\odot$. }
	\label{tab:overshoot_prescriptions}
	\end{table*}

\subsection{Stellar evolution grids}\label{ss:appendix_stellar_evolution_grids}

We constructed six stellar evolution grids that, in addition to the default one, each change a single assumption relative to our default setup (step overshooting and the \citet{pauli_new_2025} wind scheme). Namely, we constructed an exponential overshooting grid (``Exp'', using MESA's scheme with $f_{\rm ov}$ rather than $\alpha_{\rm ov}$), a grid using the Dutch wind scheme from \citet{de_jager_mass_1988} and \citet{vink_mass-loss_2001}, a grid using the  wind scheme from \citet{bjorklund_new_2023}, a grid with an initial rotation of 200~km~s$^{-1}$ at ZAMS, and a grid with MLT++ enabled \citep{paxton_modules_2013}. In each case, all other physics and numerical settings are identical to the default grid.

\subsection{Further details}\label{ss:appendix_further_details}
The nested sampling setup includes a minimum of 400 live points and convergence criteria requiring a remaining evidence fraction below 0.05 and a log-evidence uncertainty below 0.5. This configuration was found to quickly converge to robust posterior distributions.

In addition to the mass-dependent overshooting prescriptions being fit to the data, all runs also fit the background component parameter $\pi_{\rm bg}$ and a separate constant $\alpha_{\rm ov}$ to fit stars with a mass larger than $M > 40\,M_\odot$ in order to reduce the effect of these stars on our main conclusions, which only intends to investigate the mass range of 12 to 40 M$_\odot$.


In our implementation, we took special care to use mass-dependent overshooting prescriptions with the same prior volume in order to ensure that the Bayesian evidence is not affected by prior volume effects. For example, for the quadratic mass-dependence, instead of using a generic $\alpha + \beta_1 M + \beta_2 M^2$ parametrisation with somewhat arbitrary priors on $\beta_1$, and $\beta_2$, we used the left-most, right-most and middle point in the mass range of 12 to $40\,M_\odot$, with physical priors of $\alpha_{\rm ov}$ within 0 and 0.5 for all 3 points. A quadratic mass-dependence is then defined by those 3 points.
It also ensures that the prescription never goes outside of the physical prior range at any point.

\subsection{Results}\label{ss:appendix_results}
The main results of the parameter inference runs can be found in Table~\ref{tab:median_alphas}, where the median values of the posteriors are given for each combination of mass-dependent prescription and stellar evolution model variation.

\begin{figure*}[htbp]
	\centering
	\includegraphics[width=0.8\linewidth]{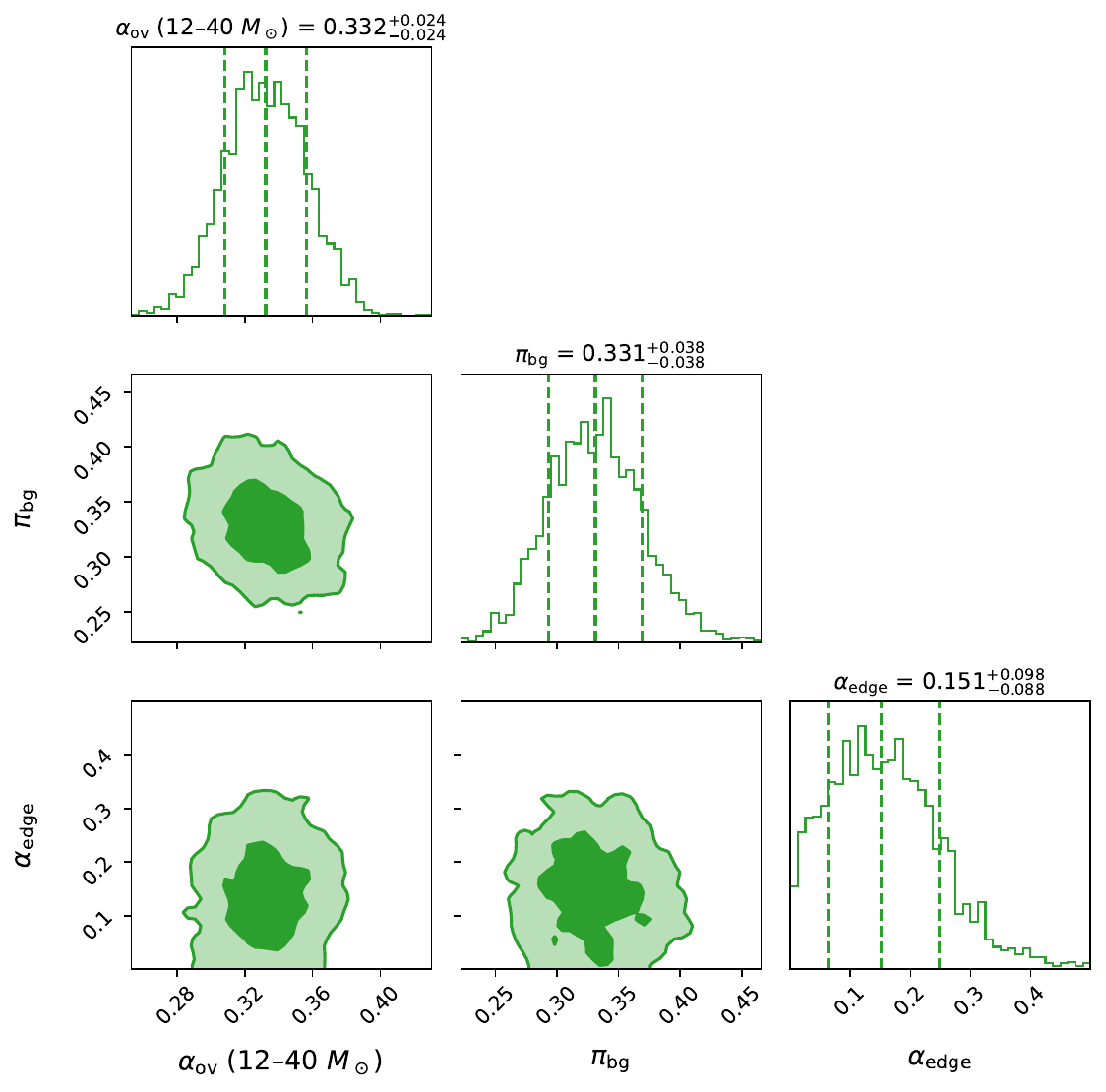}
	\caption{Corner plot of the posterior distribution for the default configuration and constant overshooting length. Alpha$_{\rm ov}$ is the overshooting length at the convective core boundary, $\pi_{\rm bg}$ is the background fraction, and $\alpha_{\rm edge}$ is the overshooting length for stars with a mass larger than 40~M$_\odot$, which is kept separate to not bias the main results.}
	\label{fig:appendix_corner_plot_default_constant}
\end{figure*}

\begin{figure*}[htbp]
	\centering
	\includegraphics[width=0.8\linewidth]{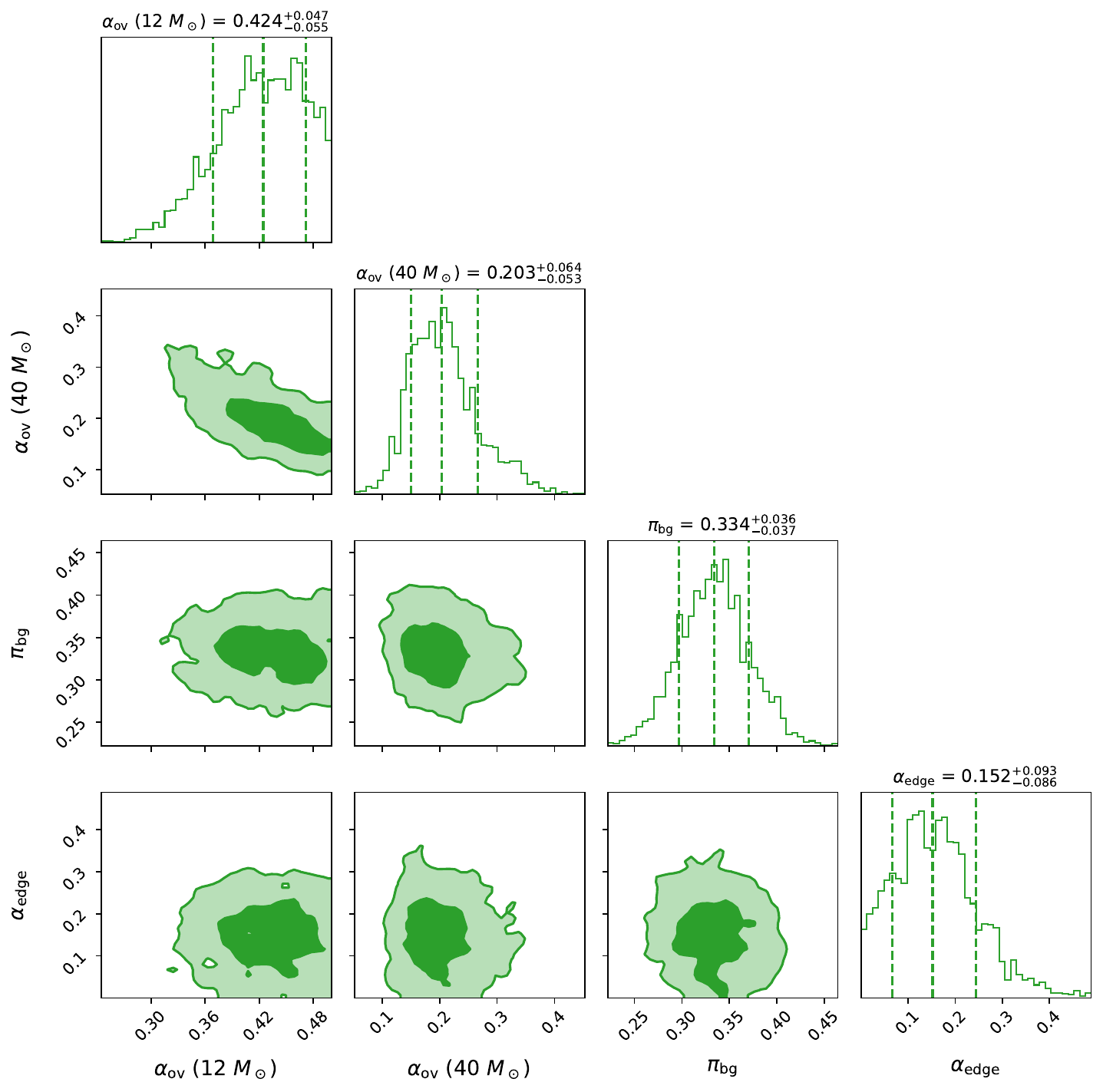}
	\caption{Corner plot of the posterior distribution for the default configuration and log-linear mass-dependence overshooting length. The main fitted parameters are the overshooting values at the boundaries of the mass range (12 and 40 M$_\odot$ respectively). For stars with masses between these boundaries, the overshooting length is log-linearly interpolated.}
	\label{fig:appendix_corner_plot_default_logm_linear}
\end{figure*}

As an example, we show a corner plot of the posterior distribution for the default configuration and constant overshooting length in Figure \ref{fig:appendix_corner_plot_default_constant}. This shows the 1D marginal posteriors for each fitted parameter, namely $\alpha_{\rm ov}$, the background fraction $\pi_{\rm bg}$, and the high-mass overshooting length that is kept separate to not influence the main results. Additionally, it shows the 2D marginal posteriors for each pair of parameters. With this model, the inferred background fraction is roughly 33\%, indicating that roughly one third of stars in this sample are not well-represented by a single-star main-sequence population. Similar background fractions are found in the other inference runs.
An example of a mass-dependent inference run is shown in Figure \ref{fig:appendix_corner_plot_default_logm_linear}, which uses a log-linear mass-dependent overshooting.

As not every inference run provided an equally good fit to the data, we also provide Bayes factors in Table~\ref{tab:bayesfactors}. For each stellar evolution model variation, the mass-dependent overshooting prescription with the highest Bayesian evidence has a Bayes factor of 1. A value higher than 1 indicates that the fit to the data is poorer, or that adding more degrees of freedom did not compensate for the relative change in data fit. A Bayes factor above 10 indicates that there is strong evidence \textit{against} this model.
From this table, we can interpret that the data has no strong discriminatory power between most prescriptions, although this depends slightly on the stellar evolution model variation.

The linear, log M linear, and exponential prescriptions are most often preferred over others, although in none of the cases are they strongly preferred over the constant prescription. If there is a mass-dependence, it seems to be slightly decreasing with mass.

\begin{figure*}[htbp]
	\centering
	\includegraphics[width=0.45\textwidth]{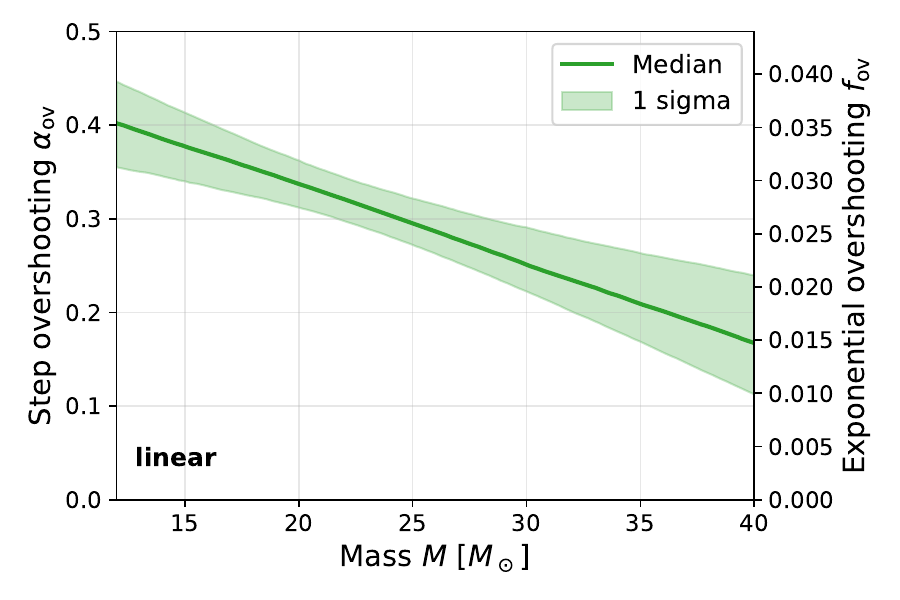}
	\includegraphics[width=0.45\textwidth]{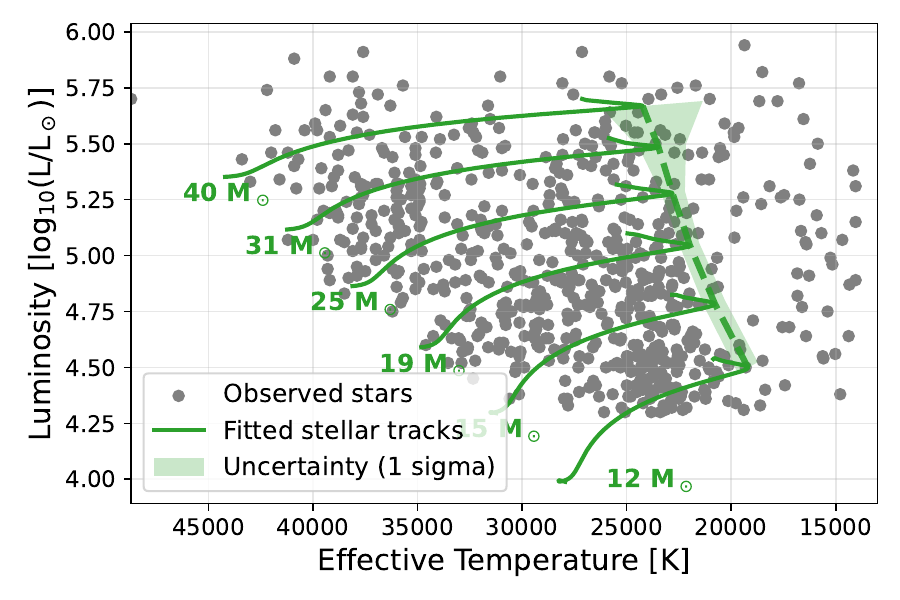}\\[0.5ex]
	\includegraphics[width=0.45\textwidth]{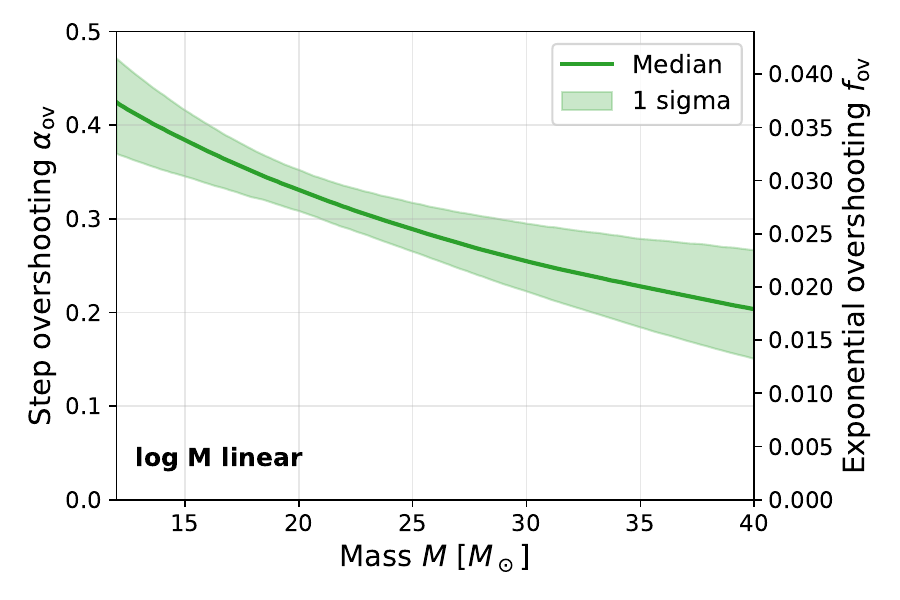}
	\includegraphics[width=0.45\textwidth]{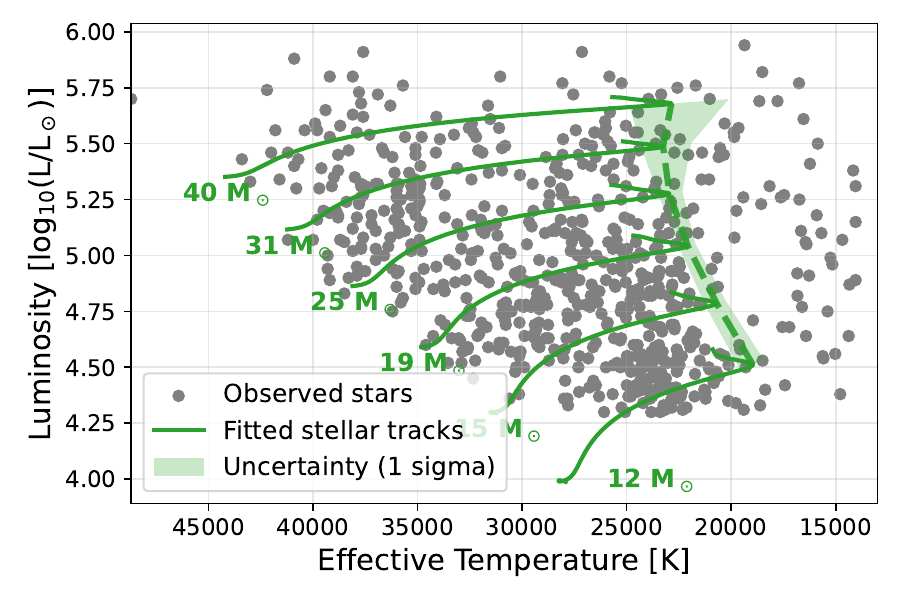}\\[0.5ex]
	\includegraphics[width=0.45\textwidth]{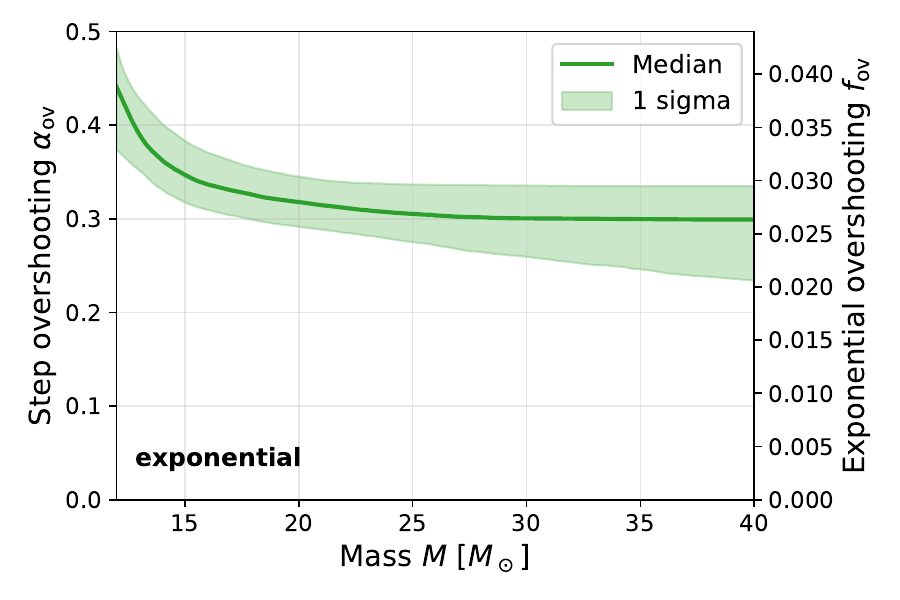}
	\includegraphics[width=0.45\textwidth]{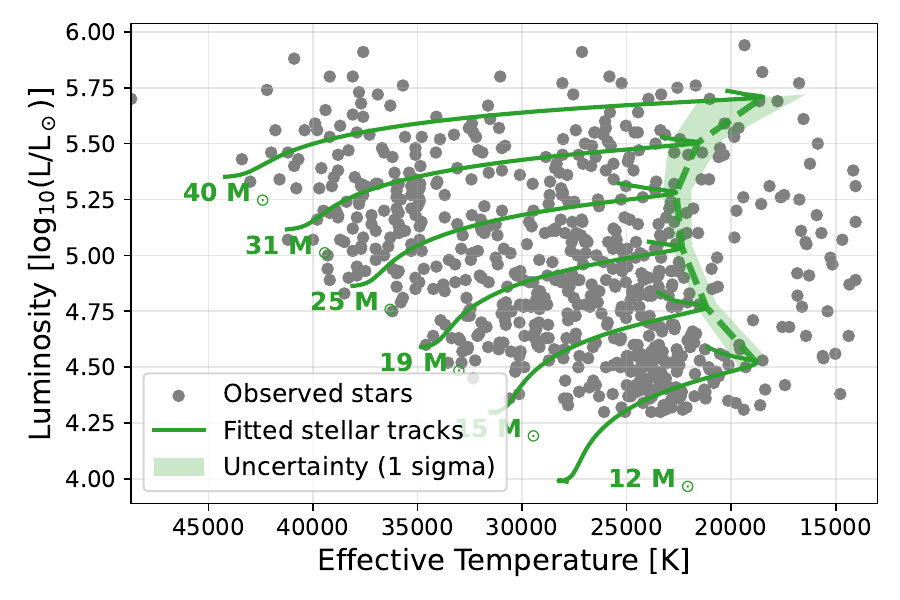}\\[0.5ex]
	\includegraphics[width=0.45\textwidth]{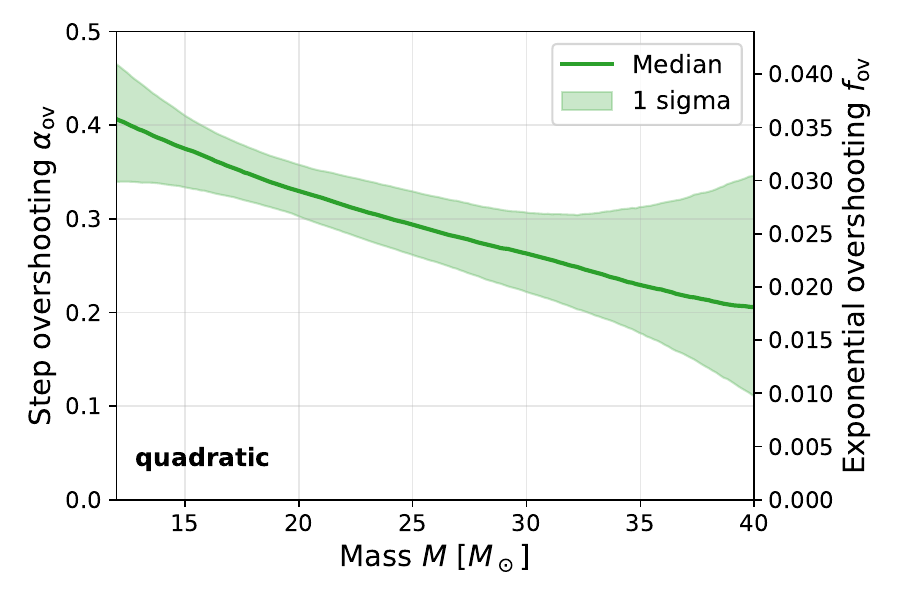}
	\includegraphics[width=0.45\textwidth]{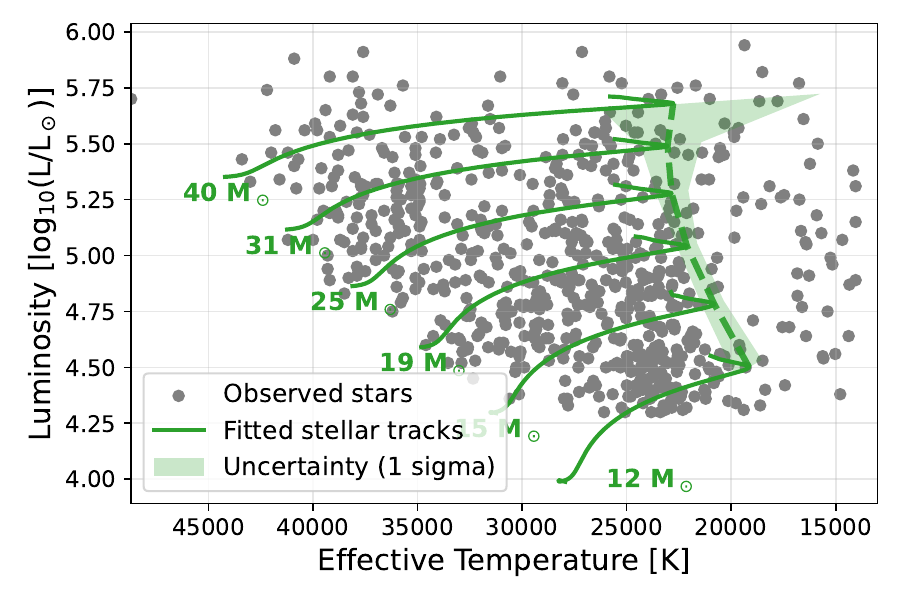}
	\caption{A comparison of the mass-dependent overshooting prescriptions for the basic configuration (page 1/2). Left column: mass-dependent overshooting. Right column: corresponding HR diagrams. Rows from top to bottom: linear, logm\_linear, exponential, and quadratic.}
	\label{fig:appendix_edge_bins_separated_basic_all_laws_1}
\end{figure*}

\begin{figure*}[htbp]
	\centering
	\includegraphics[width=0.45\textwidth]{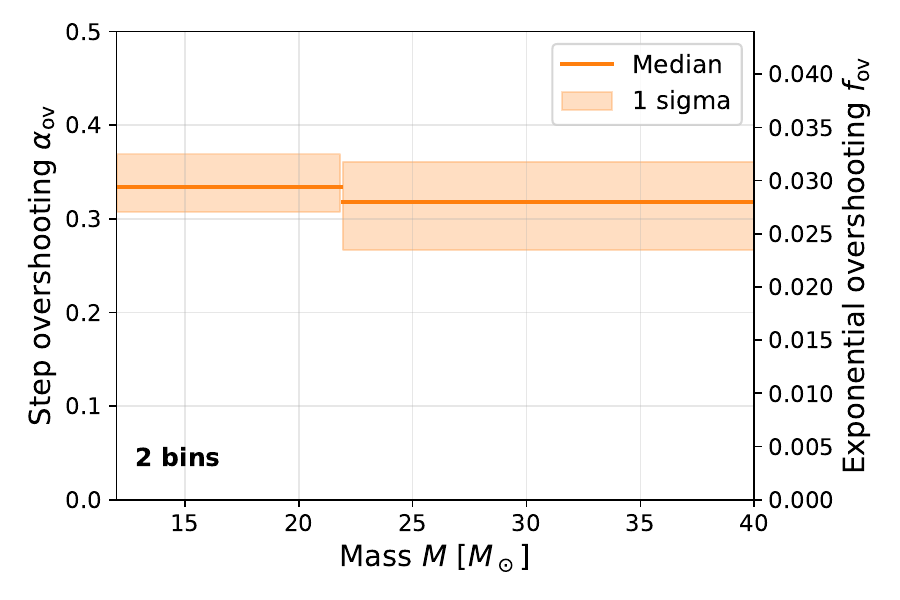}
	\includegraphics[width=0.45\textwidth]{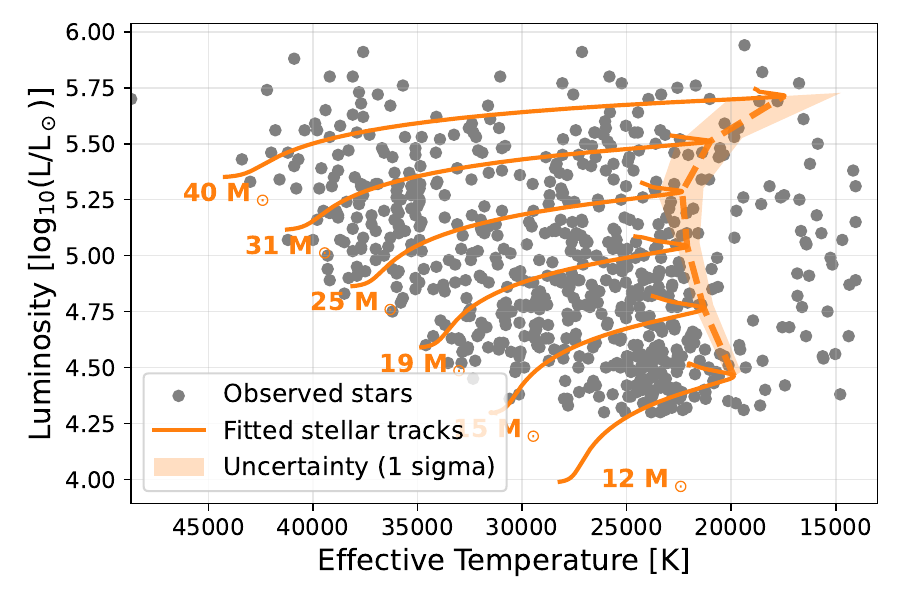}\\[0.5ex]
	\includegraphics[width=0.45\textwidth]{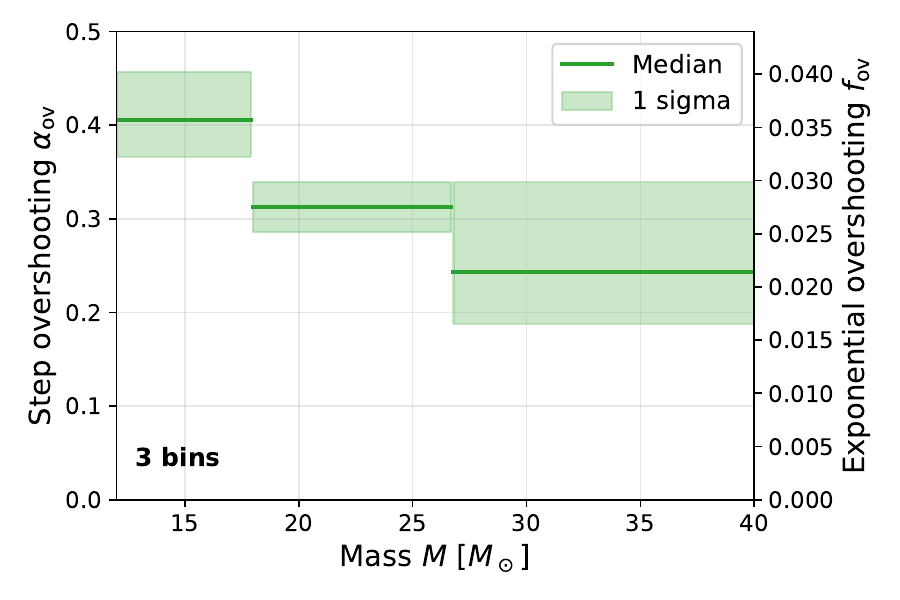}
	\includegraphics[width=0.45\textwidth]{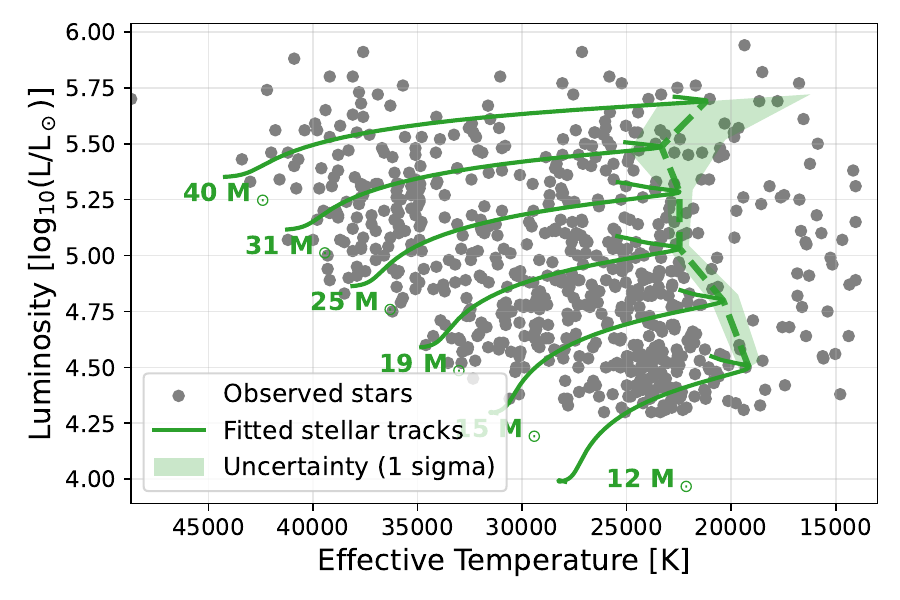}\\[0.5ex]
	\includegraphics[width=0.45\textwidth]{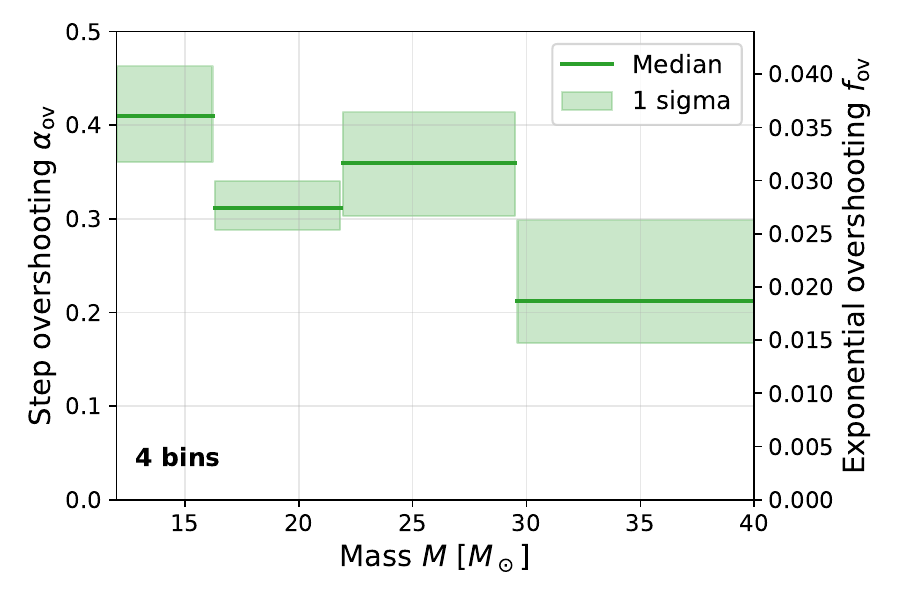}
	\includegraphics[width=0.45\textwidth]{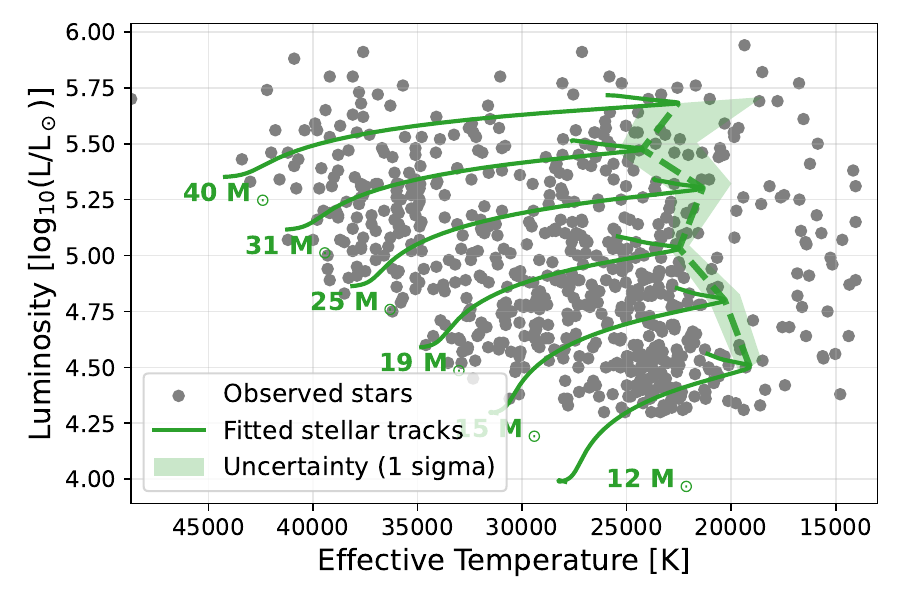}\\[0.5ex]
	\includegraphics[width=0.45\textwidth]{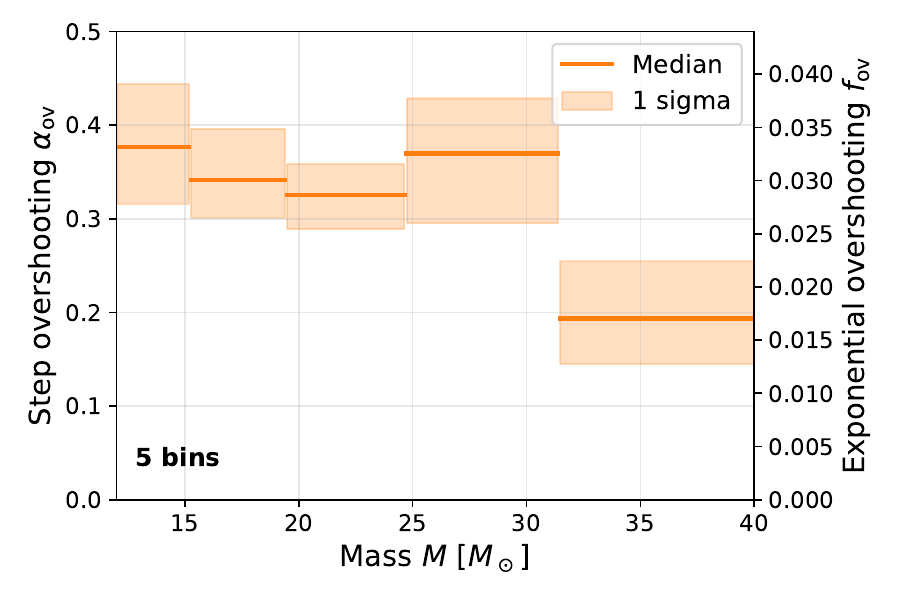}
	\includegraphics[width=0.45\textwidth]{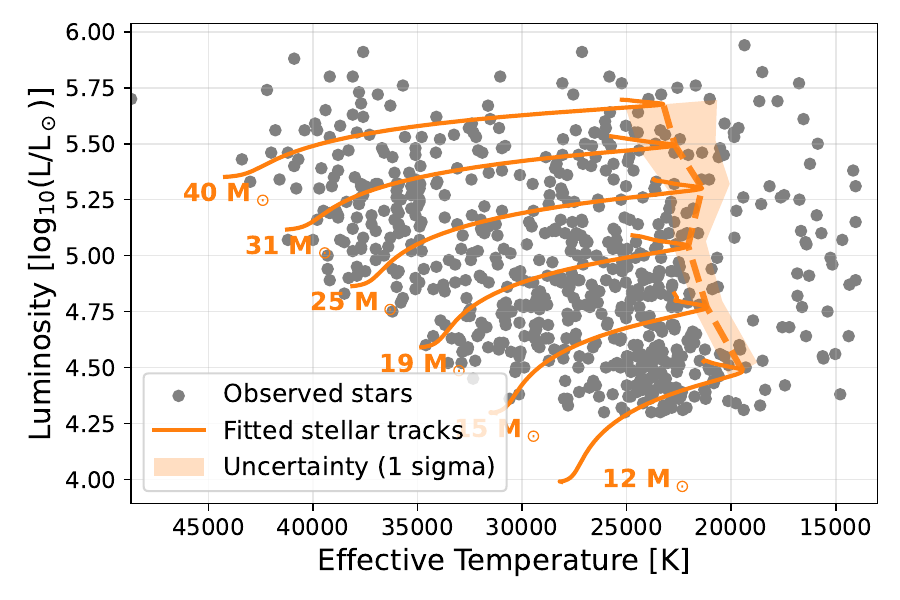}
	\caption{A comparison of the mass-dependent overshooting prescriptions for the basic configuration (page 2/2). Left column: mass-dependent overshooting. Right column: corresponding HR diagrams. Rows from top to bottom: piecewise constant with $n=2$--$5$ mass segments. Orange indicates that the prescription is strongly disfavored compared to the best fit prescription.}
	\label{fig:appendix_edge_bins_separated_basic_all_laws_2}
\end{figure*}

We also provide the main results of our default stellar models visually in Figures \ref{fig:appendix_edge_bins_separated_basic_all_laws_1} and \ref{fig:appendix_edge_bins_separated_basic_all_laws_2}, where the posteriors on the overshooting prescriptions can be seen, in addition to how this maps to the HR diagram.
These are color-coded by their Bayes factors relative to the best prescription, where green is the best prescription or equally as preferred, and orange is strongly disfavored. In the default configuration, the piecewise constant prescription with $n=2$ and $n=5$ mass segments are strongly disfavored, which likely means that the extra degrees of freedom in those mass bins do not provide a significant improvement to the fit to the data.

\begin{table*}[htbp]
	\centering
	\setlength{\tabcolsep}{1pt}
	\begin{tabularx}{\linewidth}{@{}l*{6}{>{\raggedright\arraybackslash}X}@{}}
	\toprule
	\textbf{Prescription} & \makecell[l]{Step} & \makecell[l]{Exp} & \makecell[l]{Dutch} & \makecell[l]{Björklund} & \makecell[l]{Rotating} & \makecell[l]{MLT++} \\
	\midrule
constant & \makecell[l]{$0.332$} & \makecell[l]{$0.0279$} & \makecell[l]{$0.228$} & \makecell[l]{$0.335$} & \makecell[l]{$0.334$} & \makecell[l]{$0.338$} \\
\hline
linear & \makecell[l]{$0.503$ \\ $- 0.00838\,M$} & \makecell[l]{$0.0433$ \\ $- 7.59\times 10^{-4}\,M$} & \makecell[l]{$0.353$ \\ $- 0.00620\,M$} & \makecell[l]{$0.491$ \\ $- 0.00786\,M$} & \makecell[l]{$0.377$ \\ $- 0.00336\,M$} & \makecell[l]{$0.491$ \\ $- 0.00731\,M$} \\
\hline
log $M$ linear & \makecell[l]{$0.881$ \\ $- 0.423\,\log_{10} M$} & \makecell[l]{$0.0843$ \\ $- 0.0434\,\log_{10} M$} & \makecell[l]{$0.688$ \\ $- 0.352\,\log_{10} M$} & \makecell[l]{$0.881$ \\ $- 0.422\,\log_{10} M$} & \makecell[l]{$0.522$ \\ $- 0.167\,\log_{10} M$} & \makecell[l]{$0.841$ \\ $- 0.384\,\log_{10} M$} \\
\hline
quadratic & \makecell[l]{$0.539$ \\ $- 0.0123\,M$ \\ $+ 9.77\times 10^{-5}\,M^{2}$} & \makecell[l]{$0.0530$ \\ $- 0.00160\,M$ \\ $+ 1.73\times 10^{-5}\,M^{2}$} & \makecell[l]{$0.466$ \\ $- 0.0156\,M$ \\ $+ 1.89\times 10^{-4}\,M^{2}$} & \makecell[l]{$0.535$ \\ $- 0.0119\,M$ \\ $+ 9.12\times 10^{-5}\,M^{2}$} & \makecell[l]{$0.348$ \\ $- 0.00275\,M$ \\ $+ 1.45\times 10^{-5}\,M^{2}$} & \makecell[l]{$0.509$ \\ $- 0.00903\,M$ \\ $+ 4.41\times 10^{-5}\,M^{2}$} \\
\hline
exponential & \makecell[l]{$0.299$ \\ $+ 22.0\,e^{-0.420M}$} & \makecell[l]{$0.0245$ \\ $+ 5.35\,e^{-0.478M}$} & \makecell[l]{$0.204$ \\ $+ 18.7\,e^{-0.379M}$} & \makecell[l]{$0.297$ \\ $+ 15.5\,e^{-0.389M}$} & \makecell[l]{$0.304$ \\ $+ 9.70\,e^{-0.456M}$} & \makecell[l]{$0.313$ \\ $+ 23.7\,e^{-0.436M}$} \\
\hline
$n{=}2$ & \makecell[l]{$0.334$ \\ $0.318$} & \makecell[l]{$0.0279$ \\ $0.0269$} & \makecell[l]{$0.240$ \\ $0.196$} & \makecell[l]{$0.335$ \\ $0.320$} & \makecell[l]{$0.294$ \\ $0.320$} & \makecell[l]{$0.341$ \\ $0.330$} \\
\hline
$n{=}3$ & \makecell[l]{$0.405$ \\ $0.313$ \\ $0.243$} & \makecell[l]{$0.0356$ \\ $0.0257$ \\ $0.0203$} & \makecell[l]{$0.292$ \\ $0.218$ \\ $0.156$} & \makecell[l]{$0.405$ \\ $0.312$ \\ $0.252$} & \makecell[l]{$0.301$ \\ $0.289$ \\ $0.321$} & \makecell[l]{$0.411$ \\ $0.321$ \\ $0.267$} \\
\hline
$n{=}4$ & \makecell[l]{$0.410$ \\ $0.312$ \\ $0.360$ \\ $0.212$} & \makecell[l]{$0.0364$ \\ $0.0258$ \\ $0.0317$ \\ $0.0164$} & \makecell[l]{$0.312$ \\ $0.229$ \\ $0.229$ \\ $0.143$} & \makecell[l]{$0.409$ \\ $0.314$ \\ $0.361$ \\ $0.220$} & \makecell[l]{$0.292$ \\ $0.246$ \\ $0.328$ \\ $0.264$} & \makecell[l]{$0.411$ \\ $0.319$ \\ $0.378$ \\ $0.237$} \\
\hline
$n{=}5$ & \makecell[l]{$0.376$ \\ $0.341$ \\ $0.325$ \\ $0.370$ \\ $0.193$} & \makecell[l]{$0.0327$ \\ $0.0286$ \\ $0.0272$ \\ $0.0328$ \\ $0.0145$} & \makecell[l]{$0.298$ \\ $0.254$ \\ $0.218$ \\ $0.230$ \\ $0.116$} & \makecell[l]{$0.379$ \\ $0.345$ \\ $0.324$ \\ $0.378$ \\ $0.200$} & \makecell[l]{$0.292$ \\ $0.321$ \\ $0.284$ \\ $0.364$ \\ $0.256$} & \makecell[l]{$0.378$ \\ $0.345$ \\ $0.336$ \\ $0.393$ \\ $0.222$} \\
\bottomrule
\end{tabularx}
	\caption{The main results of the inference runs, listing the median of the posteriors for $\alpha_{\mathrm{ov}}(M)$.}
	\label{tab:median_alphas}
	\end{table*}

\begin{table*}[htbp]
	\centering
	\setlength{\tabcolsep}{4pt}
	\begin{tabularx}{\linewidth}{@{}l*{6}{>{\raggedright\arraybackslash}X}@{}}
	\toprule
	\textbf{Prescription\quad\quad\quad} & Step & Exp & Dutch & Björklund & Rotating & MLT++ \\
	\midrule
	constant & 3.81 & 3.33 & 7.89 & 3.70 & 1.06 & 3.43 \\
	linear & 1.13 & 1.15 & 1.03 & \textbf{1} & \textbf{1} & 1.32 \\
	log $M$ linear & \textbf{1} & 1.09 & 1.17 & 1.03 & 1.20 & \textbf{1} \\
	quadratic & 2.85 & 2.23 & 2.87 & 2.44 & 2.25 & 3.53 \\
	exponential & 1.98 & \textbf{1} & \textbf{1} & 1.55 & 1.61 & 1.99 \\
	$n{=}2$ & 12.2 & 10.2 & 16.9 & 11.9 & 2.87 & 9.98 \\
	$n{=}3$ & 2.45 & 3.50 & 6.49 & 2.48 & 6.15 & 2.83 \\
	$n{=}4$ & 6.16 & 6.48 & 33.8 & 5.81 & 6.29 & 6.40 \\
	$n{=}5$ & 20.6 & 22.2 & 51.8 & 21.2 & 11.1 & 27.6 \\
	\bottomrule
	\end{tabularx}
	\caption{Bayes factors relative to the best prescription in each stellar grid variation.}
	\label{tab:bayesfactors}
\end{table*}


\end{appendix}




\end{document}